\documentclass[12pt,reqno]{article}

\usepackage{fancyhdr}
\usepackage{amsmath}
\usepackage{esint}
\usepackage{amsfonts}
\usepackage{amsthm}
\usepackage{amssymb}
\usepackage{amsbsy}
\usepackage{verbatim}
\usepackage{graphicx}
\usepackage{mathrsfs}
\usepackage[hmargin = .5in,vmargin=.75in]{geometry}
\usepackage[parfill]{parskip}
\usepackage{color}
\usepackage{amstext}
\usepackage{stmaryrd}
\usepackage{enumerate}

\font\msbm=msbm10

\numberwithin{equation}{section}

\theoremstyle{plain}
\newtheorem{Theorem}{Theorem}[section]
\newtheorem{lemma}[Theorem]{Lemma}
\newtheorem{corollary}[Theorem]{Corollary}

\newtheorem{proposition}[Theorem]{Proposition}

\def\mathbb#1{\hbox{\msbm{#1}}}

\newcommand{\tr}{\operatorname{Tr}}

\newcommand{\ba}{\boldsymbol{a}}
\newcommand{\bb}{\boldsymbol{b}}

\newcommand{\bd}{\boldsymbol{d}}
\newcommand{\be}{\boldsymbol{e}}
\newcommand{\bh}{\boldsymbol{h}}

\newcommand{\bp}{\boldsymbol{p}}
\newcommand{\bq}{\boldsymbol{q}}
\newcommand{\bu}{\boldsymbol{u}}
\newcommand{\bv}{\boldsymbol{v}}
\newcommand{\bw}{\boldsymbol{w}}
\newcommand{\bx}{\boldsymbol{x}}
\newcommand{\by}{\boldsymbol{y}}
\newcommand{\bz}{\boldsymbol{z}}

\newcommand{\BA}{\boldsymbol{A}}
\newcommand{\BB}{\boldsymbol{B}}
\newcommand{\BC}{\boldsymbol{C}}
\newcommand{\BD}{\boldsymbol{D}}
\newcommand{\BF}{\boldsymbol{F}}
\newcommand{\BI}{\boldsymbol{I}}
\newcommand{\BH}{\boldsymbol{H}}
\newcommand{\BS}{\boldsymbol{S}}
\newcommand{\BT}{\boldsymbol{T}}

\newcommand{\BV}{\boldsymbol{V}}
\newcommand{\BW}{\boldsymbol{W}}
\newcommand{\BX}{\boldsymbol{X}}
\newcommand{\BY}{\boldsymbol{Y}}
\newcommand{\BZ}{\boldsymbol{Z}}

\newcommand{\BPHIO}{\boldsymbol{\Phi}_{\Omega}}
\newcommand{\BPhiO}{\boldsymbol{\Phi}_{\Omega}}
\newcommand{\BPHI}{\boldsymbol{\Phi}}

\newcommand{\tOM}{\widetilde{\Omega}}
\newcommand{\talpha}{\widetilde{\alpha}}
\newcommand{\BPhi}{\boldsymbol{\Phi}}
\newcommand{\tbphi}{\widetilde{\boldsymbol{\phi}}}
\newcommand{\bphi}{\boldsymbol{\phi}}

\newcommand{\hbv}{\hat{\boldsymbol{v}}}

\newcommand{\hBX}{\hat{\boldsymbol{X}}}

\newcommand{\tba}{\widetilde{\boldsymbol{a}}}
\newcommand{\tbb}{\widetilde{\boldsymbol{b}}}
\newcommand{\bbb}{\bar{\boldsymbol{b}}}
\newcommand{\bba}{\bar{\boldsymbol{a}}}

\newcommand{\bzero}{\boldsymbol{0}}

\newcommand{\A}{\mathcal{A}}
\newcommand{\PP}{\mathcal{P}}
\newcommand{\OM}{\Omega}
\newcommand{\OMB}{\Omega^{\bot}}

\newcommand{\CC}{\mathbb{C}}
\newcommand{\CZ}{\mathcal{Z}}

\newcommand{\I}{\boldsymbol{I}}

\newcommand{\lag}{\langle}
\newcommand{\rag}{\rangle}
\newcommand{\Tr}{\text{Tr}}
\newcommand{\eps}{\epsilon}

\newcommand*\diff{\mathop{}\!\mathrm{d}}

\DeclareMathOperator{\supp}{supp}

\DeclareMathOperator{\VEC}{vec}
\DeclareMathOperator{\E}{E}
\DeclareMathOperator{\diag}{diag}
\DeclareMathOperator{\Range}{Range}
\DeclareMathOperator{\SNR}{SNR}
\DeclareMathOperator{\sgn}{sgn}
\DeclareMathOperator{\Ker}{Ker}
\DeclareMathOperator{\subjectto}{\text{subject to}}

\renewcommand{\qed}{\rule{2.5mm}{2.5mm}}

\begin{document}
\title{\bf Self-Calibration and Biconvex Compressive Sensing}

\author{Shuyang Ling and Thomas Strohmer
 \\ Department of Mathematics \\ University of 
California at Davis \\ Davis CA 95616 \\ \{syling,strohmer\}@math.ucdavis.edu}


\maketitle

\begin{abstract} 
The design of high-precision sensing devises becomes ever more difficult and expensive. 
At the same time, the need for precise calibration of these devices (ranging from tiny sensors to
space telescopes) manifests itself as a major roadblock in many
scientific and technological endeavors. To achieve optimal
performance of advanced high-performance sensors one must carefully
calibrate them, which is often difficult or even impossible to do in
practice.  In this work we bring together three seemingly unrelated concepts, namely Self-Calibration, Compressive
Sensing, and Biconvex Optimization. 
The idea behind self-calibration is to equip a hardware device with a smart
algorithm that can compensate automatically for the lack of calibration.
We show how several self-calibration problems can be treated efficiently 
within the framework of biconvex compressive sensing via a new method called SparseLift. More specifically, we
consider a linear system of equations $\by = \BD\BA\bx$, where both $\bx$ and the diagonal matrix $\BD$ 
(which models the calibration error) are unknown. By ``lifting'' this biconvex inverse problem we arrive at a convex optimization problem. 
By exploiting sparsity in the signal model, we
derive explicit theoretical guarantees under which both $\bx$ and $\BD$ can be recovered exactly, robustly,
and numerically efficiently via linear programming. 
Applications in array calibration and wireless communications are discussed and numerical simulations
are presented, confirming and complementing our theoretical analysis.
\end{abstract}

\section{Introduction}

The design of high-precision sensing devises becomes ever more difficult and expensive. 
Meanwhile, the need for precise calibration of these devices (ranging from tiny sensors to
space telescopes) manifests itself as a major roadblock in many
scientific and technological endeavors. 
To achieve optimal
performance of advanced high-performance sensors one must carefully
calibrate them, which is often difficult or even impossible to do in
practice. 
It is therefore necessary to build into the sensors and
systems the capability of {\sl self-calibration} -- using the
information collected by the system to simultaneously adjust the
calibration parameters and perform the intended function of the system
(image reconstruction, signal estimation, target detection, etc.).
This is often a challenging problem requiring advanced mathematical
solutions. Given the importance of this problem there has been a
substantial amount of prior work in this area.
Current self-calibration algorithms are based on joint estimation of the
parameters of interest and the calibration parameters using standard
techniques such as maximum likelihood estimation. These algorithms tend
to be very computationally expensive and can therefore be used only in
relatively simple situations. This has prevented their widespread applications in
advanced sensors.

In this paper we take a step toward developing a new framework for self-calibration by bringing together
three seemingly unrelated concepts, namely Self-Calibration, Compressive
Sensing, and Biconvex Optimization. By ``lifting'' the problem and exploiting sparsity in the signal model and/or 
the calibration model we express the biconvex problem of self-calibration as a convex optimization problem, 
which can be solved efficiently via linear programming. This new method, called {\em SparseLift}, ensures 
equivalence between the biconvex problem and the convex problem at the price of a modest increase in the
number of measurements compared to the already perfectly calibrated problem. SparseLift is inspired
by PhaseLift, an algorithm for phase retrieval~\cite{CSV11,candes13phase}, and by the blind deconvolution framework via convex programming
due to Ahmed, Recht, and Romberg~\cite{RechtRom12}.

Self-calibration is a field of research in its own right, and we do not attempt in this paper to develop a framework
that can deal with any kind of self-calibration problems. Instead our goal is more modest---we will consider a special, but at the same time in practice
very important, class of self-calibration problems.
To that end we first note that many self-calibration problems can be expressed as 
(either directly or after some modification such as appropriate discretization) 
\begin{equation}
\label{selfcalib}
\by = \BA(\bh) \bx + \bw,
\end{equation}
where $\by$ represents the given measurements, $\BA(\bh)$ is the system matrix depending on some unknown calibration
parameters $\bh$ and $\bx$ is the signal (function, image, ...)
one aims to recover and $\bw$ is additive noise.
As pointed out, we do not know $\BA(\bh)$ exactly due to the lack of calibration. 
Our goal is to recover $\bx$ and $\bh$ from $\by$. More precisely, we want to solve
\begin{equation}
\underset{\bh,\bx}{\min}\,\, \| \BA(\bh) \bx  - \by \|_2.
\label{bilinear0}
\end{equation}
If $\BA(\bh)$ depends linearly on $\bh$, then~\eqref{bilinear0} is a {\em bilinear}
inverse problem. In any case, the optimization problem \eqref{bilinear0}
is not convex, which makes its numerical solution rather challenging.

In many applications~\eqref{selfcalib} represents an underdetermined system even if $\BA(\bh)$ is perfectly known.
Fortunately, we can often assume $\bx$ to be sparse (perhaps after expressing it in a suitable basis,
such as a wavelet basis). Moreover, in some applications  $\bh$ is also
sparse or belongs to a low-dimensional subspace. Hence this suggests considering
\begin{equation}
\underset{\bh,\bx}{\min} \,\,\| \BA(\bh) \bx  - \by \|_2 + f(\bx,\bh),
\label{biconvex0}
\end{equation}
where $f(\bx,\bh)$ is a convex function that promotes sparsity in $\bx$
and/or $\bh$. Often it is reasonable to assume that
\eqref{biconvex0} is {\em biconvex}, i.e., it is convex in $\bh$ for
fixed $\bx$ and convex in $\bx$ for fixed $\bh$. In other cases, one can
find a close convex relaxation to the problem. We can thus frequently
interpret~\eqref{biconvex0} as a {\em biconvex compressive sensing}
problem. 

Yet, useful as~\eqref{biconvex0} may be, it is still too difficult and too general to develop
a rigorous and efficient numerical framework for it. 
Instead we focus on the important special case 
\begin{equation}
\label{basiceq}
\by = \BD(\bh) \BA \bx + \bw,
\end{equation}
where $\BD(\bh)$ is a diagonal matrix depending on the unknown parameters $\bh$ and $\BA$ is known. 
Of particular interest is the case where $\bx$ is sparse and the diagonal entries of $\BD$ belong to some known subspace.
Hence, suppose we are given
\begin{equation}\label{model}
\by = \BD\BA\bx + \bw, \quad \BD = \diag(\BB\bh),
\end{equation}
where $\by\in\CC^{L\times 1}$, $\BD:\CC^{L\times L}$, $\bh\in\CC^{k\times 1}$, $\BA\in \CC^{L\times N}$,  $\BB\in\CC^{L\times k}$ and $\bx\in\CC^{N\times 1} $ is $n$-sparse.
$\BD = \diag(\BB\bh)$ can be interpreted as the unknown calibrating parameters lying in the range of a linear
operator $\BB$. Indeed, this model will be the main focus of our investigation in this paper. More examples and
explanations of the model can be found in the next two subsections as well as in Sections~\ref{ss:arraynumerics} and \ref{ss:5Gnumerics}.

Before moving on to applications, we pause for a moment to discuss the issue of identifiability.
It is clear that if $(\BD,\bx)$ is a pair of solutions to~\eqref{basiceq}, then so is $(\alpha
\BD,\frac{1}{\alpha}\bx$) for any non-zero $\alpha$. Therefore, it is only possible to uniquely reconstruct
$\BD$ and $\bx$ up to a scalar. Fortunately for most applications this is not an issue, and we will
henceforth ignore this scalar ambiguity when we talk about recovering $\BD$ and $\bx$.

The reader may think that with the model in~\eqref{basiceq} we finally have reached a level of simplicity that is trivial
to analyze mathematically and no longer useful in practice. However, as is often the case in mathematics, a seemingly simple
model can be deceptive. On the one hand a rigorous  analysis of the  diagonal
calibration model above requires quite non-trivial mathematical tools,
and on the other hand this ``simple'' model does arise in numerous important applications. Perhaps the most widely known
instance of~\eqref{basiceq} is {\em blind deconvolution}, see e.g.~\cite{RechtRom12,LWD11}.
Other applications include radar, wireless communications, geolocation, direction finding, X-ray crystallography, cryo-electron microscopy, 
astronomy, medical imaging, etc.  We briefly discuss two such applications in more detail below. 

\subsection{Array self-calibration and direction-of-arrival estimation}
\label{ss:doa}

Consider the problem of direction-of-arrival estimation using an array
with $L$ antennas. Assuming that $n$ signals are impinging on the array,
the output vector (using the standard narrowband model) is given by~\cite{TF09}
\begin{equation}
\label{doa1}
\by(t) = \sum_{k=1}^n \BD(\bh) \ba(\theta_k,\psi_k)  x_k(t)  + \bw(t).
\end{equation}
The (unknown) calibration of the sensors is captured by the  $L \times L$ matrix $\BD$.
By discretizing the array manifold $ \ba(\theta_k,\psi_k)$, where $(\theta,\psi)$ are the azimuth and elevation
angles, respectively,  we can arrange equation~\eqref{doa1} as (see~\cite{FS14})
\begin{equation}
\label{doa2}
\by= \BD(\bh) \BA \bx + \bw,
\end{equation}
where $\bw$ now also includes discretization error.
Here $\ba(\theta,\psi)$ is a completely known function of $(\theta,\psi)$
and does not depend on calibration parameters, so $\BA$ is a known
matrix.   

To give a concrete example, assuming for simplicity all sensors and signal sources lie in the same plane, for a circular array with uniformly spaced antennas, the $k$-th column of the $L\times N$ matrix $\BA$ is given by 
$\{e^{-2\pi i \langle \bv_k , \bu_j \rangle}\}_{j=0}^{L-1}$ with
$\bv_k = \begin{bmatrix}
\sin(\theta_k) &
\cos(\theta_k)
\end{bmatrix}^T,
\bu_j = \frac{\lambda}{2 \sin(\pi/L)}
 \begin{bmatrix}
 \sin(\frac{2\pi j}{L}) &
 \cos(\frac{2\pi j}{L})
\end{bmatrix}^T,
$
where $-\frac{\pi}{2} \le \theta_0 < \dots < \theta_{N-1} \ge \frac{\pi}{2}$
is a discretization of the angular domain (the possible directions of arrival), the entries of $\bu = \{u_j\}_{j=0}^{L-1}$ represent the geometry
of the sensor array,  $L$ is the number of sensors, and $\lambda$ is the array element spacing in wavelengths.

An important special case is when $\BD(\bh)$ is a diagonal matrix whose diagonal elements represent unknown
complex gains associated with each of the antennas. The calibrating issue may come from position offsets of each antenna and/or  from gain discrepancies
caused by changes in the temperature and humidity of the environment.
If the true signal directions $(\bar{\theta}_1,\bar{\psi}_1), \dots, (\bar{\theta}_n,\bar{\psi}_n)$
lie on the  discretized grid of angles $(\theta_1, \psi_1) \cdots, (\theta_N,\psi_N)$ corresponding to the columns of $\BA$,   
 then the  $N\times 1$ vector $\bx$ is $n$-sparse, otherwise $\bx$ is only approximately $n$-sparse. 
In the direction finding problem we usually have multiple snapshots of the measurements $\by(t)$, in which case
$\by$ and $\bx$ in~\eqref{doa2} become matrices, whose  
columns are associated with different snapshots~\cite{FS14}.

\subsection{The Internet of Things, random access, and sporadic traffic}
\label{ss:5G}

The Internet of Things  (IoT) will connect billions of wireless devices, which is far more than the current Fourth-Generation (4G)  wireless system
can technically and economically accommodate. One of the many challenges of 5G, the future and yet-to-be-designed wireless communication
system, will be its ability to manage the massive number of \emph{sporadic traffic} generating IoT devices which
are most of the time inactive, but regularly access the network 
for minor updates with no human interaction~\cite{WBSJ14}. Sporadic traffic will dramatically increase in the 5G market and it is a common understanding among 
communications engineers that this traffic cannot be handled within the current 4G random access procedures. 
Dimensioning the channel access according to classical information and communication
theory results in a severe waste of resources which does not scale towards the requirements of the IoT.
This means among others that the overhead caused by the exchange of certain types of information between transmitter and receiver, such 
as channel estimation, assignment of data slots, etc, has to be avoided.  In a nutshell, if there is a huge number of devices each with the purpose to sporadically transmit
a few bits of data, we must avoid transmitting each time a signal where the part carrying
the overhead information is much longer than the part carrying the actual data.  The question is whether this is
possible. 

In mathematical terms we are dealing with the following problem (somewhat simplified, but it captures the essence).
For the purpose of exposition we consider the single user/single antenna case. Let $\bh$ denote
the channel impulse response and $\bx$ the signal to be transmitted, where $\bx$ is a sparse signal, its non-zero
coefficients containing the few data that need to be transmitted.  We encode $\bx$ by computing $\bz = \BA \bx$, 
where $\BA$ is a fixed precoding matrix, known to the transmitter and the receiver. However, the non-zero locations of $\bx$ are not
known to the receiver, and nor is $\bh$, since this is the overhead information that we want to avoid transmitting,
as it will change from transmission to transmission.  The received signal is given by the
convolution of $\bh$ and $\bz$, i.e., $\by = \bh \ast \bz$. 
The important point here is that the receiver needs to recover $\bx$ from $\by$, but does not know $\bh$ either. This type of problem is often referred to as {\em blind deconvolution}.
After applying the Fourier transform on both sides and
writing $\BD = \diag(\hat{\bh})$ (where~$\hat{}$ denotes the Fourier transform), we obtain 
$\hat{\by} = \BD \widehat{\BA \bx}$, which is of the form~\eqref{basiceq} modulo a change of notation.
A more detailed example with concrete choices for $\BA$ and $\BD$ is given in Section~\ref{ss:5Gnumerics}.
 
\subsection{Notation and outline}

Matrices are written in uppercase boldface letters (e.g.\ $\BX$), while
vectors are written in lowercase boldface letters (e.g.\ $\bx$), and scalars are written in regular font
or denoted by Greek letters.
For a matrix $\BX$ with entries $X_{ij}$ we denote $\|\BX\|_1: = \sum_{ij} |X_{ij}|$, $\|\BX\|_*$ is
the nuclear norm of $\BX$, i.e., the sum of its singular values;
$\|\BX\|_{\infty} = \max_{ij} | X_{ij}|$, $\|\BX\|$ is the operator norm of $\BX$ and $\|\BX\|_F$ is
its Frobenius norm. $\Tr(\BX)$ denotes the trace of $\BX$.
If $\bx$ is a vector, $\|\bx\|$ is its 2-norm and $\|\bx\|_0$ denotes the number
of non-zero entries in $\bx$, and $\bar{\bx}$ denotes the complex conjugate of $\bx.$

\bigskip

The remainder of the paper is organized as follows. In Section~\ref{s:art} we briefly discuss our contributions 
in relation to the state of the art in self-calibration. Section~\ref{s:maintheorems} contains the derivation of our
approach and the main theorems of this paper. Sections~\ref{s:proof} and~\ref{s:stability} are dedicated to the 
proofs of the main theorems. We present numerical simulations in Section~\ref{s:numerics} and conclude in
Section~\ref{s:conclusion} with some open problems.

\section{Related work and our contributions} \label{s:art}

Convex optimization combined with sparsity has become a powerful tool; it provides solutions to many problems which 
were believed intractable in the fields of science, engineering and technology. Two remarkable examples are 
{\em compressive sensing} and {\em matrix completion}. Compressive sensing~\cite{Don,CanTao,FouRa13}, an ingenious
way to efficiently sample sparse signals, has made great contributions to signal processing and imaging science.
Meanwhile, the study of how to recover a low-rank matrix from a seemingly incomplete set of linear observations
has attracted a lot of attention, see e.g.~\cite{CR08,CT09,RechtSIAM}. 

Besides drawing from general ideas in compressive sensing and matrix completion, the research presented in
this work is mostly influenced by papers related to PhaseLift~\cite{CESV11,CSV11,li13sparse} and in particular by the 
very inspiring paper on blind deconvolution by Ahmed, Recht, and Romberg~\cite{RechtRom12}. 
PhaseLift provides a strategy of how to reconstruct a signal $\bx$ from its quadratic measurements $y_i = |\lag
\bx, \bz_i \rag |^2$ via ``lifting'' techniques. Here the key idea is to lift a vector-valued quadratic problem
to a matrix-valued linear problem. Specifically, we need to find a rank-one positive semi-definite matrix $\BX = \bx\bx^*$ which satisfies linear measurement equations of $\BX$: $\by_i = \Tr(\BX \BZ_i ) $ where $\BZ_i = \bz_i\bz_i^*.$ 
Generally problems of this form are NP-hard. However, solving the following nuclear norm minimization yields 
exact recovery of $\bx\bx^*:$
\begin{equation}
\min \| \BX \|_*, \quad \subjectto \quad y_i = \Tr(\BX\BZ_i), \BX \succeq 0.
\label{nuclear_phase}
\end{equation}

This idea of "Lifting" also applies to blind deconvolution \cite{RechtRom12}, which refers to the recovery of two unknown
signals  from their convolution.  The model can be converted into the form \eqref{model}  via applying the Fourier 
transform and under proper assumptions $\BX_0 = \bh_0\bx_0^T$ can be recovered exactly by solving a nuclear norm minimization program in the following form,
\begin{equation}
\min \| \BX \|_*, \quad \subjectto \quad y_i = \Tr(\BX\BA_i), \quad \BA_i \text{ is a rank-one matrix},
\label{nuclear}
\end{equation}
 if the number of measurements is at least larger than the dimensions of both $\bh$ and $\bx$. Compared with PhaseLift \eqref{nuclear_phase}, the difference  is that  $\BX$ can be asymmetric and there is of course no guarantee for positivity. 
Ahmed, Recht, and Romberg derived a very appealing framework for blind deconvolution via convex programming, and we borrow
many of their ideas for our proofs.

The idea of our work is motivated by the two examples mentioned above and our results are related 
to and inspired by those of Ahmed, Recht, and Romberg. However, there are important differences in our setting. First, we consider
the case where one of the signals is sparse. It seems unnecessary to use as many measurements as the signal length 
and solve \eqref{nuclear} in that case. Indeed, as our results show we can take much fewer measurements in such a
situation. This means we end up with a system of equations that is underdetermined even in the perfectly 
calibrated case (unlike~\cite{RechtRom12}), which forces
us to exploit sparsity to make the problem solvable. Thus~\cite{RechtRom12} deals with a bilinear
optimization problem, while we end up with a biconvex optimization problem.
Second, the matrix $\BA$ 
in~\cite{RechtRom12} is restricted to be a Gaussian random matrix, whereas in our case $\BA$ can be a random 
Fourier matrix as well. This additional flexibility is relevant in real applications. For instance a practical
realization of the application in 5G wireless communications as discussed in Section~\ref{ss:5G} and
Section~\ref{ss:5Gnumerics}
would never involve a Gaussian random matrix as spreading/coding matrix, but it could definitely incorporate 
a random Fourier matrix.

Bilinear compressive sensing problems have also been investigated in e.g.~\cite{CM14ID,WJ13,LWB13}.
These papers focus on the issue of injectivity and the principle of identifiability. Except for the general setup, there is little
overlap with our work. Self-calibration algorithms have been around in the literature for quite some time, see
e.g.~\cite{FW91,See94,BN07,LLZ2011} for a small selection, with blind deconvolution being a prominent special case, cf.~\cite{LWD11} and the references
therein. It is also worth noting that~\cite{GCD12,BPGD13,SCKZ13} have developed self-calibration algorithms by using
sparsity, convex optimization, and/or ideas from compressive sensing. In particular,~\cite{GCD12, BPGD13} give many motivating examples of self-calibration and propose similar methods 
 by using the lifting trick and minimizing $\|\cdot\|_1$ or $\|\cdot\|_1 + \lambda\|\cdot\|_*$ to exploit sparsity and/or low-rank property. While these algorithms are definitely useful and interesting and give many numerical results, most of them do not provide
any guarantees of recovery. To the best of our knowledge, our paper is the first work to provide theoretical guarantee for recovery after bringing self-calibration problems into the framework of biconvex compressive sensing. 
Uncertainties in the sensing matrix have been analyzed for instance
in~\cite{HS10,CSP11}, but with the focus on quantifying the uncertainty, while our current work can be seen as a means to mitigate the uncertainty in the sensing matrix.

\section{Self-calibration, biconvex optimization, and SparseLift}
\label{s:maintheorems}


As discussed in the introduction, we are concerned with 
$\by = \BD(\bh)\BA\bx + \bw$, where the diagonal matrix $\BD$ and $\bx$ are unknown.
Clearly, without further information it is impossible to recover $\bx$ (or $\BD$) from $\by$
since the number of unknowns is larger than the number of measurements. What comes to our rescue is the
fact that in many applications additional knowledge about $\BD$ and $\bx$ is available. Of particular
interest is the case where $\bx$ is sparse and the diagonal entries of $\BD$ belong to some known subspace.

Hence, suppose we are given
\begin{equation}\label{model}
\by = \BD\BA\bx + \bw, \quad \BD = \diag(\BB\bh),
\end{equation}
where $\by\in\CC^{L\times 1}$, $\BD:\CC^{L\times L}$, $\bh\in\CC^{k\times 1}$, $\BA\in \CC^{L\times N}$,  $\BB\in\CC^{L\times k}$ and $\bx\in\CC^{N\times 1} $ is $n$-sparse.
E.g., $\BB$ may be a matrix consisting of the first $k$ columns of the DFT (Discrete Fourier Transform) matrix which
represents a scenario where the diagonal entries in $\BD$ are slowly varying from entry to entry.
One may easily formulate the following optimization problem to recover $\bh$ and $\bx$,
\begin{equation}\label{bilinear}
\underset{\bh,\bx}{\min}\| \diag(\BB\bh) \BA\bx - \by \|_2.
\end{equation}
We can even consider adding an $\ell_1$-component in order to promote sparsity in $\bx$,
\begin{equation}\label{biconvex}
\underset{\bh, \bx}{\min} \| \diag(\BB\bh) \BA\bx - \by \|_2 + \lambda \|\bx\|_1, \quad \lambda > 0.
\end{equation}

However, both programs above are nonconvex. For \eqref{bilinear}, it is an  optimization with a bilinear objective function because it is linear with respect to one vector when the other is fixed. For \eqref{biconvex}, the objective function is not even bilinear but biconvex \cite{gorski07biconvex}. 

\subsection{SparseLift}

To combat the difficulties arising from the bilinearity of the objective function in \eqref{bilinear}, we apply a 
novel idea by ``lifting'' the problem  to its
feasible set \cite{candes13phase, CSV11} . It leads to a problem of recovering a rank-one matrix $\bh\bx^T$ instead
of finding two unknown vectors. Denote by $\ba_i$ the $i$-th column of $\BA^T$ and by $\bb_i$ the $i$-th column of 
$\BB^*$. A little linear algebra yields
\begin{equation}\label{def:A}
y_i = (\BB\bh)_i \bx^T\ba_i = \bb_i^*\bh \bx^T\ba_i.
\end{equation}
Let $\quad \BX := \bh\bx^T$ and define the linear operator $\A: \CC^{k\times N}\rightarrow \CC^L$,
\begin{equation*}
\by = \A(\BX) := \{ \bb_i^* \BX \ba_i \}_{i=1}^L
\end{equation*}
Now the problem is to find a rank-one matrix $\BX$  satisfying the {\em linear} constraint $\A(\BX) = \A(\BX_0)$ where $\BX_0 = \bh_0\bx^T_0$ is the true matrix. A popular method is to take advantage of the rank-one property and use nuclear norm minimization. 
\begin{equation*}
\min \|\BX\|_*, \quad  \subjectto  \quad \A(\BX) = \A(\BX_0)
\end{equation*}
where $\|\BX\|_* $ denotes the nuclear norm of $\BX$ which is the sum of singular values.

Before we proceed, we need a bit of preparations.
By using the inner product defined on $\CC^{k\times N}$ as $\lag \BX, \BZ\rag: = \tr(\BX\BZ^*)$,  the adjoint
operator $\A^*(\bu): \CC^L \rightarrow \CC^{k\times N}$ of $\A(\BX) = \{ \bb_l^* \BX \ba_l\}_{l=1}^L$ and
$\A^*\A(\BX)$ have their forms as
\begin{equation*}
\A^*(\bu): = \sum_{l=1}^L u_l \bb_l \ba_l^*, \quad \A^*\A(\BX) = \sum_{l=1}^L \bb_l\bb_l^* \BX \ba_l\ba_l^*.
\end{equation*}

We also need a matrix representation $\BPhi: L\times kN$ of $\A: \A(\BX) = \{\bb_l^*\BX\ba_l\}_{l=1}^L$ such that
\begin{equation}\label{def:Phi}
\BPhi \VEC(\BX) = \VEC(\A(\BX)),
\end{equation}
$\VEC(\BX)$ reshapes matrix $\BX$  into a long column vector in lexicographic order.
For each $l$, $\bb_l^* \BX \ba_l = (\ba_l \otimes \bbb_l)^T \VEC(\BX)$ follows from the property of Kronecker
product, where $\bbb_l$ is the complex conjugate of $\bb_l$. It naturally leads to the block form of $\BPhi^*$,
\begin{equation}
\BPHI^* =
\left[
\bphi_1, \cdots,  \bphi_l,  \cdots, \bphi_L
\right]
_{l=1}^L,\quad \bphi_l = \bba_l\otimes \bb_l : kN\times 1.
\end{equation}

With all those notations in place, the following formula holds
\begin{equation*}
\quad \VEC(\A^*\A(\BX)) 
= \BPhi^*\BPhi \VEC(\BX) 
= \sum_{l=1}^L \bphi_l \bphi_l^* \VEC(\BX) 
= \sum_{l=1}^L \left[(\bba_l\bba_l^*)\otimes (\bb_l\bb_l^*)\right] \VEC(\BX).
\end{equation*}

$\BPhi$ also has an explicit representation of its columns.
Denote 
\begin{equation}\label{def:phist}
\tbphi_{i,j} := \diag(\tbb_i) \tba_j, \quad i = 1,2,\cdots, k \text{ and } j = 1,2,\cdots, N.  
\end{equation}
$\tbb_i$ and $\tba_j$
are the $i$-th and $j$-th column of $\BB$ and $\BA$ respectively. Both $\tbb_i$ and $\tba_j$ are  of the size
$L\times 1$. Then we can write $\BPhi$ into
 \begin{equation*}
 \BPhi = \left[ \tbphi_{1,1}, \cdots, \tbphi_{k,1}, \cdots, \tbphi_{1,N}, \cdots, \tbphi_{k,N} \right]_{kN\times
L}.
 \end{equation*}

Assume we pick a subset $\Gamma_p \subset \{1,2,\cdots, L\}$ with $|\Gamma_p| = Q$, we define $\A_p(\BX) :=  \{
\bb_l^* \BX \ba_l \}_{l\in\Gamma_p}$ from $\CC^{k\times N}\rightarrow \CC^Q$ and $\A_p^*\A_p(\BX) : =
\sum_{l\in\Gamma_p} \bb_l\bb_l^* \BX\ba_l\ba_l^*.$
$\A_p$ also has its matrix representation $\BPHI_p: Q\times kN $ as
\begin{equation}\label{PhiP}
\BPHI_p^* =
\left[
\bphi_1, \cdots,  \bphi_l,  \cdots, \bphi_L
\right]
_{l\in\Gamma_p},
\end{equation}
such that $\VEC(\A_p(\BX)) = \BPhi_p\VEC(\BX)$ and $\VEC(\A_p^*\A_p(\BX)) = \BPhi_p^*\BPhi_p\VEC(\BX).$

Concerning the support of $\bx_0$, $\BX_0$ and $\VEC(\BX_0)$, we can assume without loss of generality that the 
support of $\bx_0$ is its first $n$ entries and thus $\BX_0: = \bh_0\bx_0^T$ have its first $n$ columns nonzero. By 
a slight abuse of notation, we write $\OM$  for the supports of $\bx_0$, $\BX_0$ and $\VEC(\BX_0)$. Let $\bx_{\OM}$ 
and $\BX_{\OM}$ denote the orthogonal projections of $\bx$ and $\BX$ on $\OM$, respectively.
 Define $\A_{\OM}$ and $\A^*_{\OM}$ as the restriction of $\A$ and $\A^*$ onto $\OM$,
 \begin{equation}\label{def:AOM}
 \A_{\OM} (\BX) = \{ \bb_l^* \BX_{\OM} \ba_l\}_{l=1}^L = \{ \bb_l^*\BX_{\OM} \ba_{l,\OM} \}_{l=1}^L,\quad \A^*_{\OM}(\bu) = \sum_{l=1}^L u_l \bb_{l}\ba_{l,\OM}^* 
 \end{equation}
One can naturally define $\BPHIO: L\times kN$ with non-zero columns $\{\tbphi_{i,j}\}_{1\leq i\leq k, 1\leq j \leq n}$  satisfying $\BPhi_{\OM} \VEC(\BX) = \A_{\OM}(\BX)$ and $\BPHI_{\OMB}:L\times kN$ with non-zero columns $ \{\tbphi_{i,j}\}_{1\leq
i\leq k, j > n}$ such that $\BPhi^*_{\OMB} \bu = \VEC(\sum_{l=1}^L u_l \bb_l\ba_{l,\OMB}^*).$

Even if we stipulate that the degrees of freedom in $\BD$ is smaller than $L$ by assuming that the diagonal
entries of $\BD$ belong to some subspace, without any further knowledge about $\bx$ the self-calibration problem
is not solvable, since the number of unknowns exceeds the number of measurements.  Fortunately, in many cases we can assume that
$\bx$ is sparse. Therefore, the matrix $\BX = \bh\bx^T$ is not only of rank one but also sparse. A common method to
recover $\BX$ from $\by$ is to use a linear combination of $\|\cdot\|_* $ and $\|\cdot\|_1$ as an objective function (See \cite{li13sparse} for an example in phase retrieval) and solve the following convex program
\begin{equation}\label{l1+lambdanuc}
\min   \|\BX\|_* + \lambda\|\BX\|_1, \lambda >0, \quad \subjectto \quad   \A(\BX) = \A(\BX_0).
\end{equation}
 \eqref{l1+lambdanuc} tries to impose both sparsity and rank-one property of a matrix. While~\eqref{l1+lambdanuc}
seems a natural way to proceed,  we will
choose a somewhat different optimization problem for two reasons. First, it may not be
straightforward to find a proper value for $\lambda$, see also Figures \ref{l1+0dot1nuc} and \ref{l1+10nuc} 
in Section~\ref{s:numerics}. Second, in~\cite{OJF12} it is shown that if we perform convex optimization by
combining the two norms as in~\eqref{l1+lambdanuc}, then we can do no better, order-wise, than an algorithm
that exploits only one of the two structures.  Since only minimizing $\|\cdot\|_*$ requires $L = O(N+k)$ (which can be too large for large $N$)  and there also holds $\|\BZ\|_1 \ge \|\BZ\|_*$ for any matrix $\BZ$, 
both of them suggest that perhaps in certain cases it may suffice to simply try to minimize $\|\BX\|_1$, which
in turn may already yield a solution with sparse structure and small nuclear norm. Indeed, as well will demonstrate,  $\ell_1$-minimization performs well enough to recover $\BX_0$ exactly under certain conditions although it seemingly fails to take advantage  of the rank-one property. Hence we consider
\begin{equation}\label{l1}
\min  \|\BX\|_1,  \quad \subjectto \quad \A(\BX) = \A(\BX_0).
\end{equation}
This idea first lifts the recovery problem of two unknown vectors to a matrix-valued problem and then exploits
sparsity through $\ell_1$-minimization. We will refer to this method as \textit{SparseLift}. It takes advantage of the ``linearization'' via lifting,
while maintaining the efficiency of compressive sensing and linear programming.

If the observed data are noisy, i.e., $\by = \A(\BX_0) + \bw$ with $\|\bw\| < \eta$, we take the following alternative convex program
\begin{equation}\label{l1_noise}
\min  \|\BX\|_1, \quad \subjectto \quad \|\A(\BX) - \A(\BX_0)\| < \eta.
\end{equation}

It is useful to rewrite \eqref{l1} in matrix-vector form. Noticing that $\A(\BV) = \BPhi\bv$ for $\bv = \VEC(\BV)$, $\BV\in\CC^{k\times N}$ and $\bv\in \CC^{kN\times 1}$, we have the following equivalent formulation of \eqref{l1}
\begin{equation}\label{l1_matrix}
\min  \| \bv \|_1, \quad \subjectto \quad \BPhi\bv = \BPhi\bv_0,
\end{equation}
where $\bv_0 := \VEC(\BX_0)$. The matrix form of \eqref{l1_noise} is obtained similarly,
\begin{equation}\label{l1_noise_matrix}
\begin{array}{rl}
\min  \| \bv \|_1, \quad \subjectto \quad \|\BPhi\bv - \by\| \leq \eta.
\end{array}
\end{equation}
This model attempts to recover $\BX_0$ from noisy observation $\by$, where  $\by =  \BPhi\bv_0 +  \bw $ and $\|\bw\| \leq \eta.$

\subsection{Main results}

Our main finding is that SparseLift enables self-calibration if we are willing to take a few more additional
measurements. Furthermore, SparseLift is robust in presence of additive noise. In our theory we consider two
choices for the matrix $\BA$: in the one case $\BA$ is a Gaussian random matrix; in the other case $\BA$
consists of rows chosen uniformly at random with replacement from the $N \times N$  Discrete Fourier Transform (DFT)
matrix $\BF$ with $\BF^{\ast} \BF = \BF \BF^{\ast} = N \BI_N$. In the latter case, we simply refer to $\BA$ as a random Fourier matrix. Note that we normalize $\BF$ to $\BF^*\BF = N\I_N$ simply because $\E(\ba \ba^*) = \I_N$ where $\ba$ is a column of $\BA^*$. 
\begin{Theorem}
\label{th:main}
Consider the model \eqref{model} and assume that $\BB$ is an $L\times k$ tall matrix with $\BB^*\BB = \I_k$, $\bh$
is a $k\times 1$ dense vector and $\bx$ is an $n$-sparse and $N\times 1$ vector. Then the solution $\hat{\BX}$ to $(\ref{l1})$ is equal to  $\bh_0\bx_0^T$ with probability at least $1 - O(L^{-\alpha+1})$ if

\begin{enumerate}
\item $\BA$ is an $L\times N$ real Gaussian random matrix with $L < N$ with each entry $A_{ij} \stackrel{iid}{\sim} \mathcal{N}(0,1)$ and

\begin{equation}\label{gaussian_measure}
\frac{L}{\log^2 L} \geq C_{\alpha} \mu^2_{\max} kn \log( kN).
\end{equation}

\item $\BA$ is an $L \times N$ random Fourier matrix with $L < N$ and
\begin{equation}\label{fourier_measure}
L = PQ, \quad P \geq \frac{\log(4\sqrt{2kn}\gamma)}{\log 2}, \quad Q \geq C_{\alpha} \mu^2_{\max} kn (\log(L) + \log(kN))
\end{equation}
where $\gamma = \sqrt{2N(\log(2kN) + 1) + 1}$. 
\end{enumerate}
In both cases
$C_{\alpha} >0$ is growing linearly with $\alpha$ and  $\mu_{\max}$ is defined as the largest absolute value in
the matrix $\sqrt{L}\BB$, i.e.,
\begin{equation*}
\mu_{\max}:= \max_{ij} \sqrt{L} |B_{ij}|.
\end{equation*}
\end{Theorem}
\textbf{Remarks:} 
We note that $\max_{1\leq l\leq L}\|\bb_l\|^2 \leq \frac{k}{L} \mu^2_{\max}$, $|B_{ij}| \leq \frac{\mu_{\max}}{\sqrt{L}}$ and $\mu_{\max} \geq 1$. 
The definition of $\mu_{\max}$ seems strict but it contains a lot of useful and common examples.  For example, a matrix $\BB$ has $\mu_{\max} = 1$ if each column is picked from an $L\times L$ unitary DFT matrix and $\BB^*\BB = \I_k.$ We expect the magnitude of each entry in $\BB$ does not vary too much from entry to entry and thus the rows of $\BB$ are incoherent.  

\begin{Theorem}\label{main_thm_noise}
If the observed $\by$ is contaminated by noise, namely, $\by  = \A(\BX_0) + \bw$ with $\| \bw \| \leq \eta$, the solution $\hat{\BX}$ given by (\ref{l1_noise}) has the following error bound.
\begin{enumerate}
\item If $\BA$ is a Gaussian random matrix,
\begin{equation*}
\|\hat{\BX} - \BX_0 \|_F \leq (C_0 + C_1\sqrt{kn}) \eta
\end{equation*}
with probability of success at least $1 - O(L^{-\alpha+1})$. $L$ satisfies $(\ref{gaussian_measure})$ and both $C_0$ and $C_1$ are both constants. 
\item If $\BA$ is a random Fourier matrix,
\begin{equation*}
\|\hat{\BX} - \BX_0 \|_F \leq (C_0 + C_1\sqrt{P}\sqrt{kn}) \eta
\end{equation*}
with probability of success at least $1 - O(L^{-\alpha+1})$. $L$ and $P$ satisfy $(\ref{fourier_measure})$ and both $C_0$ and $C_1$ are both constants 
\end{enumerate}
\end{Theorem}
The proof of the theorem above is introduced in Section~\ref{s:stability}.
The solution $\hat{\BX}$  obtained by SparseLift, see \eqref{l1} or \eqref{l1_noise} for noisy data, is not necessarily
rank-one. Therefore, we compute the singular value decomposition of $\hat{\BX}$ and set $\hat{\bh} =
\sqrt{\hat{\sigma}_1} \hat{\bu}_1$ and $\hat{\bx} = \sqrt{\hat{\sigma}_1} \hat{\bv}_1$, where $\hat{\sigma}_1$ is
the leading singular value and $\hat{\bu}_1$ and $\hat{\bv}_1$ are the leading left and right singular vectors, respectively. 

\begin{corollary}\label{colvec}
Let $\hat{\sigma}_1 \hat{\bu}_1\hat{\bv}_1^T$ be the best Frobenius norm approximation to $\hat{\BX}$, then there exists a scalar $\alpha_0$ and a constant $c_0$ such that
\begin{equation*}
\| \bh_0 - \alpha_0 \hat{\bh}\| \leq c_0\min(\eps/\|\bh_0\|, \|\bh_0\|), \quad \| \bx_0 - \alpha_0^{-1} \hat{\bx}\| \leq c_0\min(\eps/\|\bx_0\|, \|\bx_0\|)
\end{equation*} 
where $\eps = \|\hat{\BX} - \BX_0\|_F.$
\end{corollary}

\cite{RechtRom12, CSV11, CL12PL} also derived error bounds of recovering the signals from the recovered "lifted" matrix by using the leading eigenvector or leading left and right singular vectors. \cite{CSV11,CL12PL} focus on PhaseLift and the recovered vector from $\hat{\BX}$ is unique up to a unit-normed scalar. Our corollary is more similar to Corollary 1 in \cite{RechtRom12} because both of them deal with a bilinear system instead of a quadratic system like PhaseLift. There are two signals to recover from $\hat{\BX}$ in our case and thus they are unique up to a scalar $\alpha_0$.  Without prior information, no theoretical estimate can be given on how large $\alpha_0$ is. 

The proof of Corollary \ref{colvec} consists of considering the leading
eigenvalue and eigenvector of the Hermitian matrices $\hat{\BX}^*\hat{\BX}$ and $\hat{\BX}\hat{\BX}^*$ and then using the  
$\sin$-$\theta$ Theorem in \cite{davis70rotation}. The same idea is implemented in the proof of Theorem 1.2  from \cite{CSV11} where the detailed reasoning can be found.

We could use a mixed $\ell_2$-$\ell_1$-norm in~\eqref{l1} to enforce column-sparsity in $\BX$, i.e., replace $\|\BX\|_1$ by $\|X\|_{2,1}$. Indeed, analogous versions of our main results,
Theorem~\ref{th:main} and Theorem~\ref{main_thm_noise}, hold for the mixed $\ell_2$-$\ell_1$-norm case as well, which will be reported in a forthcoming paper.

\section{Proof of Theorem~\ref{th:main}}
\label{s:proof}

The outline of the proof of Theorem~\ref{th:main} follows a meanwhile well-established pattern in the literature
of compressive sensing~\cite{candes_plan2011,FouRa13} and low-rank matrix recovery, but as is so often the case, the difficulty lies in the 
technical details.  We first derive sufficient conditions for exact recovery.
The fulfillment of these conditions involves computing norms of certain random matrices as well as the
construction of a dual certificate. The main part of the proof is dedicated to these two steps.
While the overall path of the proof is similar for the case when $\BA$ is a Gaussian random matrix or a random Fourier
matrix, there are several individual steps that are quite different. 

\subsection{A sufficient condition for exact recovery}

We will first give a sufficient condition for exact recovery by using SparseLift. Then we prove all the proposed assumptions are valid. 

\begin{proposition}
The rank-1 matrix $\BX_0 = \bh_0\bx_0^T$ with support $\Omega$ is the unique minimizer to SparseLift \eqref{l1} if 
\begin{equation}\label{suff_cond1}
\lag \sgn(\BX_0), \BH \rag + \|\BH_{\Omega^{\bot}}\|_1 > 0 
\end{equation}
for any $\BH\in\Ker(\A)\backslash\{0\}$.
\end{proposition}
This proposition can be regarded as a sufficient condition of exact recovery of a sparse vector via $\ell_1$ minimization in compressive sensing (See Theorem 4.26(a) in \cite{FouRa13}. They are actually the same but appear in different formulations).
Based on (\ref{suff_cond1}), we propose an alternative sufficient condition which is much easier to use. 

\begin{proposition}
$\BX_0$ is the unique minimizer  if there exists a $\BY\in\text{Range}(\A^*)$ satisfying the following conditions, 
\begin{equation}
\label{cond2}  \|\sgn(\BX_0) - \BY_{\Omega}\|_F \leq \frac{1}{4\sqrt{2}\gamma}, \quad \|\BY_{\Omega^{\bot}}\|_{\infty} \leq \frac{1}{2} 
\end{equation}
and the linear operator $\A$ satisfies 
\begin{equation}
\label{cond1} \| \A^*_{\Omega} \A_{\Omega} - \I_{\OM}\|\leq \frac{1}{2}, \quad \|\A\| \leq \gamma.
\end{equation}
\end{proposition}
\textbf{Remark: } We will later specify $\gamma$ in \eqref{upboundgs} and \eqref{upbound}.

\begin{proof}
The proof starts with $(\ref{suff_cond1})$. Note that $\lag \sgn(\BX_0), \BH\rag = \lag \sgn(\BX_0) - \BY,\BH\rag$ if $\BY\in\Range(\A^*)$ and $\BH\in\Ker(\A)\backslash\{0\}$. By splitting the inner product of $\lag \sgn(\BX_0) - \BY, \BH\rag$ on $\OM$ and $\OMB$, the following expression can guarantee the exact recovery of $l_1$ minimization,
\begin{equation*}
\lag \sgn(\BX_0) - \BY_{\Omega}, \BH_{\Omega}\rag - \lag  \BY_{\Omega^{\bot}}, \BH_{\Omega^{\bot}}\rag +
\|\BH_{\Omega^{\bot}}\|_1 >0 .
\end{equation*}

By using the H\"{o}lder inequality, we arrive at the stronger condition \begin{equation}\label{suff_cond3}
-\|\sgn(\BX_0) - \BY_{\Omega}\|_F \|\BH_{\Omega}\|_F + (1 - \|\BY_{\Omega^{\bot}}\|_{\infty})
\|\BH_{\Omega^{\bot}}\|_1 > 0 .
\end{equation}

Due to $ \| \A^*_{\Omega} \A_{\Omega} - \I_{\OM}\|\leq \frac{1}{2}$ and $\|\A\| \leq \gamma$, we have $\|\A(\BH_{\Omega})\|_F \geq \frac{1}{\sqrt{2}} \|\BH_{\Omega}\|_F$ and $\|\A(\BH_{\Omega^{\bot}})\|_F \leq \gamma  \|\BH_{\Omega^{\bot}}\|_F $. As a result, 
\begin{equation}\label{A_rip}
\frac{1}{\sqrt{2}}  \|\BH_{\Omega}\|_F \leq \|\A(\BH_{\OM})\|_F = \|\A(\BH_{\OMB})\|_F \leq \gamma
\|\BH_{\Omega^{\bot}}\|_F \leq \gamma  \|\BH_{\Omega^{\bot}}\|_1.
\end{equation}

The second equality follows from $\A(\BH_{\OM}) + \A(\BH_{\OMB}) = \bzero$. Substitute (\ref{A_rip}) into (\ref{suff_cond3}) and we only need to prove the following expression in order to achieve exact recovery,
\begin{equation*}
\left(1 - \|\BY_{\Omega^{\bot}}\|_{\infty} - \sqrt{2}\gamma \|\sgn(\BX_0) - \BY_{\Omega}\|_F\right) \|\BH_{\Omega^{\bot}}\|_1 > 0.
\end{equation*}
 If $\|\sgn(\BX_0) - \BY_{\Omega}\|_F \leq \frac{1}{4\sqrt{2}\gamma},  \|\BY_{\Omega^{\bot}}\|_{\infty} \leq \frac{1}{2}$ and  $\BH_{\Omega^{\bot}} \neq 0$, the above expression 
is strictly positive. On the other hand, $\BH_{\Omega^{\bot}} = 0$ implies $\BH= 0$ from \eqref{A_rip}. Therefore,
the solution $\hat{\BX}$ of SparseLift equals  $\BX_0$ if the proposed conditions  \eqref{cond1} and \eqref{cond2} are satisfied. 

\end{proof}

\subsection{Local isometry property}

In this section, we will prove Lemma~\ref{weakrip}, the local isometry property,  which achieves the first condition in \eqref{cond1}. 

\begin{lemma}\label{weakrip}
For linear operators $\A$ and $\BPhi$ defined in~\eqref{def:A} and~\eqref{def:Phi} respectively,
\begin{equation*}
\|\BPhi^*_{\OM} \BPhi_{\OM} - \I_{\OM}\| = \|\A_{\OM}^*\A_{\OM} - \I_{\OM}\| \leq \delta 
\end{equation*}
holds true with probability at least $1 - L^{-\alpha +1}$ where $\I_{\OM}$ is the identity operator on $\OM$ if 
\begin{enumerate}
\item  $\ba_l$ is the $l$-th row of an $L\times N$ Gaussian random matrix $\BA$ and $L \geq C_{\alpha} \mu^2_{\max}kn \max\{ \log L /\delta^2, \log^2L /\delta\}$.
\item $\ba_l$ is the $l$-th row of an $L\times N$ random Fourier matrix $\BA$ and $L\geq C_{\alpha} \mu^2_{\max}kn \log L/\delta^2$.
\end{enumerate}
In both cases, $C_{\alpha}$ grows linearly with respect to $\alpha$ and $\alpha > 1.$
\end{lemma} 

In fact, it suffices to prove  Lemma \ref{lowbound} and Lemma \ref{keylemma} because 
Lemma \ref{weakrip} is a special case of Lemma \ref{keylemma} when $Q = L$ and $P=1$.

\subsubsection{Two ingredients of the proof}
Before we move to the proof of  Lemma \ref{lowbound} and Lemma \ref{keylemma}, we introduce two important ingredients which will be used later.
One ingredient concerns the rows $\{\bb_l\}_{l=1}^L$ of $\BB$. There exists a disjoint partition of index set $\{1,2,\cdots, L\}$ into $\{ \Gamma_p \}_{p=1}^{P'}$ with each $|\Gamma_p| = Q'$  and $Q' > C \mu^2_{\max} k\log(L)$ such that
\begin{equation}\label{propsuppB1}
\max_{1\leq p\leq P}\left\| \BT_p - \frac{Q'}{L}\I_k \right\| < \frac{Q'}{4L},  \text{ where }\BT_p = \sum_{l\in\Gamma_p} \bb_l\bb_l^*
\end{equation}
The existence of such a partition follows from the proof of Theorem 1.2 in \cite{candes07sparsity} and is used in Section 3.2 of \cite{RechtRom12}.
\eqref{propsuppB1} also can give a simple bound for the operator norms of  both $ \BT_p$ and $ \BT_p^{-1}.$ 
\begin{equation}\label{propsuppB}
\max_{1\leq p\leq P}\left\| \BT_p \right\| \leq \frac{5Q'}{4L} ,
\quad\quad 
\max_{1\leq p\leq P}\left\| \BT_p^{-1} \right\| \leq \frac{4L}{3Q'}
\end{equation}

The other ingredient concerns the non-commutative Bernstein inequality. For Gaussian and Fourier cases, the inequalities are slightly different. This following theorem, mostly due to Theorem 6.1 in \cite{tropp12user},  is used for proving the Fourier case of \eqref{low_st}.
\begin{Theorem}\label{BernFourier}
Consider a finite sequence of $\CZ_l$ of independent centered random matrices with dimension $M\times M$. Assume that $\|\CZ_l \| \leq R$ and introduce the random matrix 
\begin{equation}\label{S}
\BS = \sum_{l\in\Gamma_p} \CZ_l.
\end{equation}
Compute the variance parameter
\begin{equation}\label{sigmasq}
\sigma^2 = \max\Big\{ \| \sum_{l=1}^Q \E(\CZ_l\CZ_l^*)\|, \| \sum_{l=1}^Q \E(\CZ_l^* \CZ_l)\| \Big\},
\end{equation}
then for all $t \geq 0$
\begin{equation*}
\Pr( \|\BS\| \geq t) \leq 2M\exp \left( -\frac{t^2/2}{ \sigma^2 + Rt/3}\right).
\end{equation*}
\end{Theorem}
For the Gaussian case we replace $\|\CZ_l \| \leq R$ with $\|\CZ_l\|_{\psi_1} \leq R$ where the norm $\|\cdot\|_{\psi_1}$ of a matrix is defined as
\begin{equation*}
\|\BZ\|_{\psi_1} := \inf_{u \geq 0} \{ \E[ \exp(\|\BZ\|/u)] \leq 2 \}.
\end{equation*}

The concentration inequality is different  if $\ba_l$ is Gaussian because of the unboundedness of $\BS$. We are using the form in  \cite{KolVal11}.
\begin{Theorem}\label{BernGaussian}
For a finite sequence of independent $M\times M$ random matrices $\CZ_l$ with $R : = \max_{1\leq l\leq Q}
\|\CZ_l\|_{\psi_1}$ and $\BS$ and $\sigma^2$ as defined in \eqref{S} and \eqref{sigmasq}, we have the tail bound on
the operator norm of $\BS$, 
\begin{equation}\label{gaussian_inq}
\Pr(\|\BS\| \geq t) \leq 2M \exp\left( - \frac{1}{C_0} \frac{t^2}{\sigma^2 + \log \left(
\frac{\sqrt{Q}R}{\sigma^2}\right)  Rt} \right),
\end{equation}
where $C_0$ is an absolute constant. 
\end{Theorem}

\subsubsection{A key lemma and its proof}
For each subset of indices $\Gamma_p$, we denote 
\begin{equation}\label{def:APOM}
\A_{p,\OM}(\BX) = \{ \bb_l^*\BX \ba_{l,\OM}\}_{l\in\Gamma_p}, \quad \A^*_{p,\OM}(\bu) = \sum_{l\in\Gamma_p} u_l \bb_l\ba_{l,\OM}^*, \quad \A^*_{p,\OM}\A_{p,\OM}(\BX) = \sum_{l\in\Gamma_p} \bb_l\bb_l^* \BX \ba_{l,\OM}\ba_{l,\OM}^* 
\end{equation}
Under certain conditions, it is proven that $\A^*_{p,\OM}\A_{p,\OM}$ does not deviate from the identity much for each $p$. We write this result into Lemma~\ref{lowbound}.
\begin{lemma}\label{lowbound}
For any equal partition $\{\Gamma_p \}_{p=1}^P$ of $\{1,2,\cdots, L\}$ with $|\Gamma_p| = Q$ such that
\eqref{propsuppB1}  holds true (e.g. let $Q' = Q$), we have
\begin{equation}\label{low_st}
\max_{1\leq p\leq P}\sup_{\|\BX_{\OM}\|_F = 1}\left\|  \A^*_{p,\OM}\A_{p,\OM}(\BX) - \frac{Q}{L}\BX_{\Omega}\right\|_F
 \leq \frac{Q}{2L}, 
\end{equation}
  with probability at least $1 - PL^{-\alpha} \geq 1 - L^{-\alpha + 1}$ if     \begin{enumerate}
\item  $\ba_l$ is the $l$-th row of an $L\times N$ Gaussian random matrix $\BA$ and $Q \geq C_{\alpha}\mu^2_{\max}kn \log^2 L $.
\item $\ba_l$ is the $l$-th row of an $L\times N$ random Fourier matrix $\BA$ and $Q\geq C_{\alpha} \mu^2_{\max} kn \log L$.
  \end{enumerate}
In both cases $C_{\alpha}$ grows linearly with respect to $\alpha$ and $\alpha > 1$.

\end{lemma}

 First, we prove  the following key lemma. Lemma \ref{lowbound} is a direct result from Lemma~\ref{keylemma} and the property 
stated in~\eqref{propsuppB1}. We postpone the proof of Lemma~\ref{lowbound} to Section~\ref{s:lowbound}.

\begin{lemma}\label{keylemma}
For any fixed $0 < \delta \leq 1$ and equal partition $\{\Gamma_p \}_{p=1}^P$ of $\{1,2,\cdots, L\}$ with $|\Gamma_p| = Q$ satisfying \eqref{propsuppB1} and~\eqref{propsuppB}(e.g. let $Q' = Q$), we have
\begin{equation}\label{low_st}
\max_{1\leq p\leq P}\sup_{\|\BX_{\OM}\|_F = 1}\left\|  \A^*_{p,\OM}\A_{p,\OM}(\BX) - \BT_p \BX_{\Omega}\right\|_F
 \leq \frac{\delta Q}{L}, \quad \BT_p = \sum_{l\in\Gamma_p} \bb_l\bb_l^*,
\end{equation}
  with probability at least $1 - PL^{-\alpha} \geq 1 - L^{-\alpha + 1}$ if 
  \begin{enumerate}
\item  $\ba_l$ is the $l$-th row of an $L\times N$ Gaussian random matrix $\BA$ and $Q \geq C_{\alpha} \mu^2_{\max}kn \max\{ \log L /\delta^2, \log^2L /\delta\}$. 
\item $\ba_l$ is the $l$-th row of an $L\times N$ random Fourier matrix $\BA$ and $Q\geq C_{\alpha} \mu^2_{\max}kn \log L/\delta^2$.
  \end{enumerate}
In both cases $C_{\alpha}$ grows linearly with respect to $\alpha$ and $\alpha > 1$.
\end{lemma}

\begin{proof} 
We prove the Gaussian case first; the proof of the Fourier case is similar. Let $\ba_l$ be a standard
Gaussian random vector. $\ba_{l,\OM}$ denotes the projected vector $\ba_l$ on the support of $\bx_0$ with
$|\supp(\bx_0)| = n$. We first rewrite the linear operator $\A^*_{p,\OM}\A_{p,\OM}(\BX) - \BT_p\BX_{\OM}$ into matrix form. Using the Kronecker product we define $\CZ_l$ as
follows:
\begin{equation}\label{cz}
\CZ_l := (\ba_{l,\OM}\ba_{l,\OM}^* - \I_{N,\OM})\otimes (\bb_l\bb_l^*).
\end{equation}
Every $\CZ_l$ is a centered random matrix since $\E(\ba_{l,\OM}\ba_{l,\OM}^*) = \I_{N,\OM}$, which follows from
restricting $\ba_l\ba_l^*$ to the support $\OM$ and taking its expectation. $\I_{N,\OM}$ is an $N\times N$ matrix with $\I_n$ as its $n\times n$ leading submatrix.   It is easy to check that 
\begin{equation*}
\VEC(\A^*_{p,\OM}\A_{p,\OM}(\BX) - \BT_p\BX_{\OM}) = \sum_{l\in\Gamma_p}
\CZ_l \VEC(\BX_{\OM}). 
\end{equation*}
More precisely, $\ba_{l,\OM}$ in $\CZ_l$ needs to be replaced by $\bba_{l,\OM}$ according to the property of Kronecker product. But there is no problem here since each $\ba_l$ is a real random vector. For simplicity, we denote $\CZ_l(\BX) := \bb_l\bb_l^*\BX (\ba_{l,\OM}\ba_{l,\OM}^*  - \I_{N,\OM})$. In order to use Theorem \ref{BernGaussian} to estimate $\|\sum_{l=1}^L \CZ_l \|$, we need to know $\|\CZ_l\|_{\psi_1}$ and $\sigma^2$.
We first give an estimate for the operator norm of $\CZ_l$.
\begin{equation}\label{cznorm}
\|\CZ_l\|  =  |\bb^*_{l}  \bb_l | \|\ba_{l,\Omega}\ba^*_{l,\Omega} - \I_{N,\Omega}\|  \leq  
\frac{\mu^2_{\max}k}{L}\| \ba_{l,\Omega}\ba^*_{l,\Omega} - \I_{N,\Omega} \| \leq  \frac{\mu^2_{\max}k}{L}\max(
\|\ba_{l,\Omega}\|^2, 1) .
\end{equation}

Here $\|\ba_{l,\OM}\|^2$ is a $\chi^2$ random variable with $n$ degrees of freedom. This implies that $\|\CZ_l\|_{\psi_1} \leq  \frac{C \mu^2_{\max}kn}{L}$, following from Lemma 6 in \cite{RechtRom12}. Therefore there exists an absolute constant $C_b$ such that
\begin{equation*}
R =  \max_{1\leq l\leq L} \|\CZ_l\|_{\psi_1} \leq \frac{C_b \mu^2_{\max} kn}{L}.
\end{equation*}

The second step is to estimate $\sigma^2$. It suffices to consider $\CZ_l^*\CZ_l(\BX)$ because $\CZ_l$ is a Hermitian matrix. 
\begin{equation*}
\CZ^*_l\CZ_l(\BX)  =  \|\bb_l\|^2   \bb_l\bb_l^*    \BX (\ba_{l,\Omega}\ba_{l,\Omega}^* - \I_{N,\Omega})^2.
\end{equation*}
The expectation of $\CZ_l^*\CZ_l (\BX)$ is
\begin{equation}\label{cov}
\E(\CZ_l^* \CZ_l(\BX)) = (n+1) \|\bb_l\|^2   \bb_l\bb_l^*    \BX \I_{N, \OM},
\end{equation}
following from $\E(\ba_{l,\Omega}\ba_{l,\Omega}^* - \I_{N,\Omega})^2 = (n +1)\I_{N,\OM}$ which can be proven by using similar reasoning of Lemma 9 in~\cite{RechtRom12}.
\begin{eqnarray}
\label{sigmasq} \sigma^2 & = & \left\| \sum_{l\in\Gamma_p} \E(\CZ_l^* \CZ_l) \right\| = (n+1)\left\| \sum_{l\in\Gamma_p} \|\bb_l\|^2\bb_l\bb_l^* \right\|  \\
& \leq  & \frac{\mu^2_{\max}k(n+1)}{L}\left\|   \sum_{l\in\Gamma_p} \bb_l\bb_l^*   \right\| \\
\label{sigmasq_est} & \leq &  \frac{5\mu^2_{\max}Qk(n+1)}{4L^2} \leq \frac{5\mu^2_{\max}Qkn}{2L^2} 
\end{eqnarray}
where the inequality in the second line follows from $\|\bb_l\|^2 \leq \frac{\mu^2_{\max}k}{L}$ and the positivity of $\bb_l\bb_l^*$. The only thing to do is to estimate $\log\left(\frac{\sqrt{Q}R}{\sigma^2}\right)$ before we can apply the concentration inequality \eqref{gaussian_inq}. We claim that there holds
\begin{equation*}
\log\left( \frac{\sqrt{Q}R}{\sigma^2}\right) \leq C_1 \log L
\end{equation*}
for some constant $C_1$. Notice that $Q\leq L$ and $R$ is at most of order $L$.
 It suffices to show the lower bound of $\sigma^2$ in order to prove the expression above.  Following from~\eqref{sigmasq}, we have $\sigma^2\geq (n+1)\max_{l\in\Gamma_p} \|\bb_l\|^4.$ On the other hand, ~\eqref{propsuppB1} leads to $\sum_{l\in\Gamma_p} \|\bb_l\|^2 \geq \|\sum_{l\in\Gamma_p} \bb_l\bb_l^*\| \geq \frac{3Q}{4L}$ since the trace of a positive semi-definite matrix must be greater than its operator norm. Therefore we conclude that $\sigma^2 \geq \frac{9(n+1)}{16L^2}$ because  $\max_{l\in\Gamma_p} \|\bb_l\|^2 \geq \frac{1}{Q} \sum_{l\in\Gamma_p} \|\bb_l\|^2\geq \frac{3}{4L}$ and   $|\Gamma_p| = Q.$

Now we are well prepared to use \eqref{gaussian_inq}. $\CZ_l$ can be simply treated as a $kn\times kn$ matrix for each $l$ because its support is only $kn\times kn.$  With $R\leq C_b\mu^2_{\max}kn/L$, $\sigma^2\leq \frac{5\mu^2_{\max}Qn}{2L^2}$ and $\log\left(\sqrt{Q}R/\sigma^2\right)\leq C_1\log L,$ we have
\begin{eqnarray*}
\Pr \left(\| \sum_{l\in\Gamma_p} \CZ_l \| \geq  \frac{\delta Q}{L} \right) & \leq & 2kn \exp\left( -\frac{Q^2}{C_0 L^2} \cdot \frac{\delta^2}{\frac{5\mu^2_{\max}Qkn}{2L^2} + C_b  \frac{\mu^2_{\max}kn}{L}  \cdot \frac{\delta Q}{L} \log L} \right) \\
& \leq & 2kn \exp\left( -\frac{1}{C_0} \cdot \frac{Q\delta^2}{5\mu^2_{\max}kn/2 + C_b \delta \mu^2_{\max} kn\log L}
\right) .
\end{eqnarray*}

Now we let  $Q \geq C_{\alpha} \mu^2_{\max}kn \max\{ \log L /\delta^2, \log^2L /\delta\}$ where  $C_{\alpha}$ is a function which grows linearly with respect to $\alpha$ and has sufficiently large derivative. By properly assuming $L > kn$,  $Q$ is also greater than $C_0 \mu^2_{\max}kn ( \frac{5}{2} + C_b \delta \log L) ( \alpha \log L + \log 2kn ) /\delta^2$ and

\begin{equation*}
\Pr\left( \| \sum_{l\in\Gamma_p} \CZ_l \| \geq \frac{\delta Q}{L}\right)  \leq L^{-\alpha}.
\end{equation*}
We take the union bound over all $p = 1,2,\cdots, P$ and have
\begin{equation*}
\Pr \left( \max_{1\leq p \leq P}  \| \sum_{l\in\Gamma_p} \CZ_l \| \leq \frac{\delta Q}{L}   \right) \geq 1 - P
L^{-\alpha} \geq 1 - L^{-\alpha + 1}.
\end{equation*}
This completes the proof for the Gaussian case.

We proceed to the Fourier case. 
The proof is very similar and the key is to use Theorem \ref{BernFourier}. Now we assume $\ba_l$ is the $l$-th row of a random Fourier matrix $\BA$ and thus $\E(\ba_l\ba_l^*) = \I_N$. Here $\CZ_l = (\bba_{l,\OM}\bba_{l,\OM}^* - \I_{N,\OM})\otimes \bb_l\bb_l^*$ because $\ba_l$ is a complex random vector. However, it does not change the proof at all except the notation. 
Like the proof of the Gaussian case, we need to know  $R = \max_{l\in\Gamma_p}\|\CZ_l\| $ and also $\sigma^2$. 
Following the same calculations that yielded \eqref{cznorm}, \eqref{cov} and \eqref{sigmasq_est},  we have
\begin{equation*}
R  \leq |\bb_l^*\bb_l| \max(\|\bba_{l,\OM}\|^2, 1) \leq  \frac{\mu^2_{\max}kn}{L}
\end{equation*}
and 
\begin{equation*}
\sigma^2 \leq \frac{\mu^2_{\max}k}{L}\| \sum_{l\in\Gamma_p} \bb_l\bb_l^* \BX \E( \bba_{l,\OM}\bba_{l,\OM}^* - \I_{N,\OM} )^2 \| 
\leq \frac{\mu^2_{\max}kn}{L} \|\BT_p\| 
\leq \frac{5\mu^2_{\max}Qkn}{4L^2}.
\end{equation*}
Then we apply Theorem \ref{BernFourier} and  obtain
\begin{equation*}
\Pr \left(\| \sum_{l\in\Gamma_p} \CZ_l \| \geq  \frac{\delta Q}{L} \right) \leq 2kn \exp\left( - \frac{Q\delta^2}{5\mu^2_{\max}kn/2 + 2\delta\mu^2_{\max} kn/3} \right) \leq L^{-\alpha}
\end{equation*}
if we pick $Q \geq C_{\alpha} \mu^2_{\max}kn \log L/\delta^2 $. This implies
\begin{equation*}
\Pr \left( \max_{1\leq p \leq P}  \| \sum_{l\in\Gamma_p} \CZ_l \| \leq \frac{\delta Q}{L}   \right) \geq 1 - P
L^{-\alpha} \geq 1 - L^{-\alpha + 1}.
\end{equation*}
\end{proof}

\subsubsection{Proof of Lemma \ref{lowbound}}\label{s:lowbound}
\begin{proof}
From Lemma \ref{keylemma}, we know that 
for a fixed equal partition $\{\Gamma_p \}_{p=1}^P$ of $\{1,2,\cdots, L\}$ with $|\Gamma_p| = Q$ and satisfying~\eqref{propsuppB1}, we have
\begin{equation}\label{low_st}
\max_{1\leq p\leq P}\sup_{\|\BX_{\OM}\|_F = 1}\left\| \sum_{l\in\Gamma_p} \bb_l\bb_l^* \BX_{\Omega} \ba_{l,\Omega}\ba^*_{l,\Omega} - \BT_p \BX_{\Omega}\right\|_F
 \leq \frac{\delta Q}{L}
\end{equation}
By setting $\delta  = \frac{1}{4}$ and following from $\|\sum_{l\in\Gamma_p} \bb_l\bb_l^* - \frac{Q}{L}\| \leq \frac{Q}{4L}$ in~\eqref{propsuppB1}, we have
\begin{eqnarray*}
\sup_{\|\BX_{\OM}\|_F = 1}\left\| \sum_{l\in\Gamma_p} \bb_l\bb_l^* \BX_{\Omega} \ba_{l,\Omega}\ba^*_{l,\Omega} - \frac{Q}{L} \BX_{\Omega}\right\|_F & \leq & \sup_{\|\BX_{\OM}\|_F = 1}\left\| \sum_{l\in\Gamma_p} \bb_l\bb_l^* \BX_{\Omega} \ba_{l,\Omega}\ba^*_{l,\Omega} - \BT_p \BX_{\Omega}\right\|_F \\
& & + \sup_{\|\BX_{\OM}\|_F = 1}\left\| \BT_p \BX_{\Omega} - \frac{Q}{L} \BX_{\Omega}\right\|_F \\
& \leq & \frac{Q}{4L} + \frac{Q}{4L} = \frac{Q}{2L}.
\end{eqnarray*}
\end{proof}

\subsubsection{Estimating the operator norm of $\A$}
The operator norm $\gamma$ of $\A$ with $\BA$ will be given for both the Gaussian case and the Fourier case such that $\|\A(\BX)\|_F \leq \gamma \|\BX\|_F$ for any $\BX$ with high probability. 
For the Gaussian case, Lemma 1 in \cite{RechtRom12} gives the answer directly. 
\begin{lemma}
For $\A$ defined in~\eqref{def:A} with $\BA$ a Gaussian random matrix and fixed $\alpha \geq 1$, 
\begin{equation}\label{upboundgs}
\|\A\| \leq \sqrt{N\log(NL/2) + \alpha \log L}
\end{equation}
with probability at least $1 - L^{-\alpha}.$
\end{lemma}

For the Fourier case, a direct proof will be given by using the non-commutative Bernstein inequality in Theorem~\ref{BernFourier}. 
\begin{lemma}
For $\A$ defined in~\eqref{def:A} with $\BA$ a random Fourier matrix and fixed $\alpha \geq 1$,
\begin{equation}\label{upbound}
\|\A\| \leq \sqrt{2N(\log(2kN) + 1) + 1} 
\end{equation}
with probability at least $1 - L^{-\alpha}$ if we choose $L \geq \alpha \mu^2_{\max}k \log(L).$

\end{lemma}

\begin{proof}

First note that $\|\A^*\A\| = \|\A\|^2$.
Therefore it suffices to estimate the norm of $\A^*\A,$
\begin{equation*}
\A^*\A(\BX) = \sum_{l=1}^L \bb_l\bb_l^*\BX\ba_l\ba_l^* = \sum_{l=1}^L \bb_l\bb_l^*\BX (\ba_l\ba_l^* - \I_N) + \BX.
\end{equation*}
The second equality follows from the fact that $\sum_{l=1}^L \bb_l\bb_l^* = \I_k$. 
For each centered random matrix $\CZ_l : = (\bba_l\bba_l^* - \I_N) \otimes \bb_l\bb_l^*,$ we have $\VEC(\A^*\A(\BX)) = \sum_{l=1}^L \CZ_l \VEC(\BX) + \VEC(\BX)$ and 
\begin{equation*}
\|\CZ_l\| = \|(\bba_l\bba_l^* - \I_N) \otimes \bb_l\bb_l^*\| = \|\bba_l\bba_l^* - \I_N \| \cdot \| \bb_l\bb_l^* \| \leq
\frac{\mu^2_{\max}kN}{L}.
\end{equation*}
That means $R = \max_{1\leq l\leq L} \|(\bba_l\bba_l^* - \I_N) \otimes \bb_l\bb_l^*\|  \leq \frac{\mu^2_{\max}kN}{L}$.
In order to estimate the variance of each $\CZ_l$, we only have to compute $\E(\CZ_l\CZ_l^*)$ since $\CZ_l$ is a Hermitian matrix.
\begin{equation*}
\E(\CZ_l\CZ_l^*) \leq \frac{k}{L} \E\left[  (\bba_l\bba_l^* - \I_N)^2 \otimes \bb_l\bb_l^* \right] =
\frac{\mu^2_{\max}k(N-1)}{L} \I_N\otimes \bb_l\bb_l^*.
\end{equation*}

Hence the variance of $\sum_{l=1}^L \CZ_l$ is 
\begin{equation*}
\sigma^2 \leq  \frac{\mu^2_{\max}k(N-1)}{L} \left\| \sum_{l=1}^L \I_N \otimes \bb_l\bb_l^* \right\| =
\frac{\mu^2_{\max}k(N-1)}{L} \leq \frac{\mu^2_{\max}kN}{L}.
\end{equation*}

Now we apply Bernstein's inequality and for any $t\geq 0$, we have
\begin{eqnarray*}
\Pr\left( \| \sum_{l=1}^L \CZ_l \| \geq t \right) & \leq & 2kN \cdot \exp\left( - \frac{t^2/2}{\sigma^2 + Rt/3} \right) \\
& = & 2kN \cdot \exp\left( - \frac{t^2/2}{\mu^2_{\max}kN/L + \mu^2_{\max}kNt/3L} \right).
\end{eqnarray*}

Suppose $t \geq 3$, then
\begin{eqnarray*}
\Pr\left( \| \sum_{l=1}^L \CZ_l \| \geq t \right) & \leq & 2kN \cdot \exp\left( - \frac{t^2/2}{2\mu^2_{\max}kNt/3L} \right) \\
& = & 2kN \cdot \exp\left( - \frac{3Lt}{4\mu^2_{\max}kN} \right).
\end{eqnarray*}
If we pick $t = 2N ( \log(2kN) + 1) $, then
\begin{equation*}
\Pr\left( \| \sum_{l=1}^L \CZ_l \| \geq t \right) \leq \exp\left(-\frac{L}{\mu^2_{\max}k} \right).
\end{equation*}
If $L \geq \alpha \mu^2_{\max} k \log(L)$,  we have 
\begin{equation*}
\| \A^*\A \| \leq \|\sum_{l=1}^L \CZ_l \| + 1\leq 2N (\log(2kN) + 1) + 1 \Longrightarrow \gamma: = \|\A\| \leq \sqrt{2N(\log(2kN) + 1) + 1}
\end{equation*}
with probability at least $1 - L^{-\alpha}$.
\end{proof}

\subsection{Construction of the dual certificate}

In the last section we have shown that \eqref{cond1} holds true.  Now we will prove that a $\BY\in \Range(\A^*)$
satisfying \eqref{cond2} exists by explicitly constructing it for both, the Gaussian and the Fourier case. 

\subsubsection{$\BA$ is a Gaussian random matrix}

\noindent
{\bf Construction of Exact dual certificate via least square method}: \\
It follows from Lemma \ref{weakrip} and \ref{lowbound} that $\BPHIO^*\BPHI_{\Omega}$ is invertible on $\OM$ with
$\|(\BPHIO^*\BPHIO)^{-1} \| \leq 2$ if $L > C_{\alpha}kn \log^2 L$ for some scalar $C_{\alpha}$. We construct the exact dual
certificate via (see~\cite{candes11simple})
\begin{equation}\label{gaussian_dual}
\bp  :=  \BPHI_{\OM} ( \BPHIO^*\BPHIO)^{-1}\VEC( \sgn(\BX_0)) 
, \quad \bq : =  \BPhi^* \bp = \VEC(\A^*(\bp)).
\end{equation}
Let $\BY = \A^*(\bp)$; it suffices to consider $\bq$ and verify whether $\bq_{\OM} = \VEC(\sgn(\BX_0))$ and $\|\bq_{\OMB}\|_{\infty} \leq \frac{1}{2}$.

\begin{Theorem}\label{gaussian_dual_inf_norm}
Conditioned on the event $\{ \|(\BPHIO^*\BPHIO)^{-1} \| \leq 2 \}$, there holds\begin{equation*}
\|\bq_{\OMB}\|_{\infty} \leq \frac{1}{2}
\end{equation*}
with probability at least $1 - L^{-\alpha}$ if  we pick $L > 16 \alpha \mu^2_{\max}kn ( \log(L) + \log(kN))$.
\end{Theorem}
\begin{proof}
Note $\BPHI_{\OMB}^*\bp = (\tbphi^*_{s,t} \bp)_{1\leq s\leq k, t> n}$ and $\tbphi_{s,t} = \diag(\tbb_s)\tba_t$ is independent of $\bp$ for $t > n$ because $\bp$ only uses $\tba_{j}: 1\leq j\leq n.$ In this case,
\begin{equation*}
\tbphi^*_{s,t}\bp = \tba_t^* \diag(\tbb_s^*) \bp 
\end{equation*}
where $\tba_t$ is an  $L\times 1$ standard Gaussian random vector. Due to the independence between $\diag(\tbb_s^*)\bp$ and $\ba_t$, 
\begin{equation*}
\tbphi^*_{s,t}\bp \sim \mathcal{N}(0, \|\diag(\tbb_s^*) \bp\|^2),
\end{equation*}
where
\begin{eqnarray*}
\|\diag(\tbb_s^*) \bp\|^2 & \leq & \frac{\mu^2_{\max}}{L} \|\bp\|^2 \leq \frac{\mu^2_{\max}}{L}  \VEC( \sgn( \BX_0))^* (\BPHIO^*\BPHIO)^{-1} \VEC( \sgn( \BX_0) ) \\
& \leq & \frac{2\mu^2_{\max}}{L}  \| \sgn(\BX_0)\|_F^2   = \frac{2\mu^2_{\max}kn}{L}.
\end{eqnarray*}

Therefore, $\tbphi_{s,t}^*\bp $ is a centered Gaussian random variable with variance $\sigma^2$  at most $\frac{
2\mu^2_{\max} kn }{L}$. For any $x >0$, we have
\begin{eqnarray*}
\Pr\left(|\tbphi^*_{s,t} \bp| >  x \right) & = & \frac{2}{\sqrt{2\pi \sigma^2} }\int_x^{\infty} e^{-\frac{u^2}{2\sigma^2}} \diff u  \leq   \frac{2}{\sqrt{2\pi} \sigma x} \int_x^{\infty} u e^{-\frac{u^2}{2\sigma^2}} \diff u \\
& = & -\frac{2\sigma}{\sqrt{2\pi} x} e^{-\frac{u^2}{2\sigma^2}} \Big|_x^{\infty}  =  \sqrt{\frac{2}{\pi}}
\frac{\sigma}{x} e^{-\frac{x^2}{2\sigma^2}}.
\end{eqnarray*}
Taking the union bound over all pairs $(s,t)$ of indices $(\tbphi^*_{s,t})$ satisfying $t > n$ gives
\begin{equation*}
\Pr\left(\max_{1\leq s\leq k, t>n}|\tbphi^*_{s,t} \bp| > x \right)  \leq kN  \sqrt{\frac{2}{\pi}} \frac{\sigma}{x}
e^{-\frac{x^2}{2\sigma^2}} \leq  \frac{kN \sigma}{x} e^{-\frac{x^2}{2\sigma^2}} .
\end{equation*}
If we pick $x = \frac{1}{2}$, then $\frac{\sigma}{x} \leq 2\sqrt{\frac{2\mu^2_{\max}kn}{L}}$ and
\begin{equation*}
\Pr\left(\max_{1\leq s\leq k, t>n}|\tbphi^*_{s,t} \bp| > \frac{1}{2} \right)   \leq   \frac{kN\sigma}{x}
e^{-\frac{x^2}{2\sigma^2}}  \leq  \frac{2\sqrt{2\mu^2_{\max}kn}kN}{\sqrt{L}}e^{-\frac{L}{16\mu^2_{\max}kn}} .
\end{equation*} 
Choosing $L > 16 \alpha \mu^2_{\max}nk ( \log(L) + \log(kN)) $ implies that the probability of $\{ \max_{\{1\leq s\leq k, t>n\} } |\tbphi_{s,t}^* \bp| > \frac{1}{2} \}$ will be less than $L^{-\alpha}.$
\end{proof}

\subsubsection{$\BA$ is a random Fourier matrix}
{\bf Construction of Inexact dual certificate via golfing scheme}: \\
An inexact dual certificate $\BY$ will be constructed to satisfy the conditions given in \eqref{cond2} via the
golfing scheme, which was first proposed  in \cite{gross11recovering} and has  been widely
used since then, see e.g.~\cite{RechtRom12, FouRa13, li13sparse}. The idea is to use part of the measurements in each step and construct
a sequence of random matrices $\BY_p$ recursively in $\Range(\A^*)$. The sequence converges  to $\sgn(\BX_0)$
exponentially fast and at the same time keeps each entry of $\BY_p$ small on $\OMB$. We  initialize  $\BY_0 : = \bzero$ and set
\begin{equation}\label{golfingscheme}
\BY_p := \BY_{p-1} + \frac{L}{Q} \A_p^*\A_p (  \sgn(\BX_0) - \BY_{p-1, \OM}).
\end{equation}
We denote $\BY := \BY_P$. Let $\BW_p := \BY_{p, \OM} - \sgn(\BX_0)$ be the residual between  $\sgn(\BX_0)$ and $\BY_p$
on $\OM$  and $\BW_0: = \BY_0 - \sgn(\BX_0)=  -\sgn(\BX_0)$. $\BW_p$ yields the following relations due to (\ref{golfingscheme}),
\begin{eqnarray*}
\BW_p & =  & \BW_{p-1} - \frac{L}{Q}\A_{p,\OM}^*\A_{p,\OM}(\BW_{p-1}) \\
& = & \frac{L}{Q} \left( \frac{Q}{L} - \A_{p,\OM}^*\A_{p,\OM} \right) (\BW_{p-1}).
\end{eqnarray*}
where $\A_{p,\OM}$ is defined in~\eqref{def:APOM}.
Following from Lemma \ref{lowbound}, $\|\BW_p\|_F$ decreases exponentially fast with respect to $p$ with probability at least $1 - L^{-\alpha +1}$ and
\begin{equation}\label{wnorm}
\|\BW_p\|_F \leq \frac{1}{2} \|\BW_{p-1}\|_{F}, \quad \|\BW_p\|_F \leq 2^{-p}\sqrt{kn}.
\end{equation}

The value of $P$ can be determined to make $\|\BW_P\|_F = \|\BY_{\OM} - \sgn(\BX_0)\|_F \leq \frac{1}{4\sqrt{2}
\gamma}$. Hence,
\begin{equation}\label{PP}
2^{-P} \sqrt{kn} \leq \frac{1}{4\sqrt{2}\gamma} \Longrightarrow P \geq\frac{\log(4\sqrt{2kn}\gamma)}{\log 2}.
\end{equation}

The next step is to bound the infinity norm of $\BY$ on $\OMB$. We first rewrite the recursive relation
(\ref{golfingscheme}); $\BY$ has the following form,
\begin{equation}\label{dual_cert_w}
\BY = - \frac{L}{Q}\sum_{p=1}^P \A^*_p\A_p \BW_{p-1}.
\end{equation}

Our aim is to achieve $\| \BY_{\OMB} \|_{\infty} \leq \frac{1}{2}$. It suffices to make $\| \PP_{\OMB }\A_p^*\A_p( \BW_{p-1}) \|_{\infty} \leq 2^{-p-1}\frac{Q}{L}$ in order to make $\|\BY\|_{\infty} \leq \frac{1}{2}.$ $\PP_{\OMB}$ is the  operator which orthogonally projects a matrix onto $\OMB$. For each $p$, 
\begin{equation*}
\E(\A_p^*\A_p (\BW_{p-1}) | \BW_{p-1}) = \BT_p \BW_{p-1},
\end{equation*}
since $\BW_{p-1}$ is known at step $p$ and $\A_p^*\A_p$ is independent of $\BW_{p-1}$. $\BT_p \BW_{p-1}$ has the same support as $\BW_{p-1}$ does and thus
this gives $\E \left[ \PP_{\OMB} \A_p^*\A_p(\BW_{p-1})\right] = \bzero.$
The next theorem proves that $\| \PP_{\OMB} \A_p^*\A_p(\BW_{p-1}) \|_{\infty} \leq 2^{-p-1}\frac{Q}{L}$ for all $1\leq p\leq P$ with probability at least $1 - L^{-\alpha + 1}$ if $L \geq P Q$ and $Q\geq C_{\alpha}\mu^2_{\max} kn (\log(kN) + \log(L)).$

\begin{Theorem}
Conditioned on \eqref{wnorm} and for fixed $p$,
\begin{equation*}
\Pr\left(  \| \PP_{\OMB}\A_p^*\A_p(\BW_{p-1})\|_{\infty} > \frac{Q}{2^{p+1}L} \right) \leq 2kN \exp \left(
-\frac{3Q}{128\mu^2_{\max}kn} \right).
\end{equation*} 
In particular this gives
\begin{equation*}
\Pr\left(  \| \PP_{\OMB}\A_p^*\A_p(\BW_{p-1})\|_{\infty} \leq \frac{Q}{2^{p+1}L}, \quad \forall\,\, 1\leq p\leq P \right) \geq 1 - PL^{-\alpha} \geq 1 - L^{-\alpha + 1} ,
\end{equation*}
by choosing $Q\geq \frac{128\mu^2_{\max}\alpha}{3} kn( \log(kN) + \log(L) )$ and taking the union bound for all $p$.
\end{Theorem}

\begin{proof}
For any $\bz = \VEC(\BZ)$ where $\BZ:k\times N$ has support $\OM$, we have
\begin{equation*}
\A_p^*\A_p (\BZ) = \BPHI_p^*\BPHI_p \bz
\end{equation*}
directly following from  \eqref{def:A} and \eqref{def:Phi}.
Therefore, 
\begin{equation*}
\|\PP_{\OMB}\A_p^*\A_p(\BZ)\|_{\infty} = \max_{s\in\OMB} |\lag \be_s, \BPHI_p^*\BPHI_p\bz \rag|,
\end{equation*}
where $\be_s$ is a $kN\times 1$ unit vector with its $s$-th entry equal to $1$. Here $s > kn$ because only the first $kn$ entries of $\bz$ are nonzero. The main idea of the  proof is to
estimate $|\lag \be_s, \BPHI_p^*\BPHI_p\bz\rag|$ for each $s\in\OMB$ and take the union bound over all $s\in\OMB.$
For any fixed $s\in\OMB,$
\begin{equation*}
\lag \be_s, \BPHI_p^*\BPHI_p\bz \rag =\lag \be_s, \sum_{l\in \Gamma_p} \bphi_l \bphi_l^* \bz \rag =
\sum_{l\in\Gamma_p} \lag \be_s, \bphi_l\bphi_l^* \bz \rag, \quad \bphi_l = \bba_l \otimes \bb_l.
\end{equation*}
Let $z_l := \lag \be_s, \bphi_l\bphi_l^* \bz \rag$. Notice $\E ( z_l)  = 0$ because
\begin{equation*}
\E(\lag  \be_s, \bphi_l \bphi_l^* \bz \rag ) =\be_s^* (\I_k \otimes \bb_l\bb_l^*)\bz =\be^*_s
\VEC(\bb_l\bb_l^*\BZ).
\end{equation*}
$\bb_l\bb_l^*\BZ$ and $\BZ$ have the same support and thus any entry of $\bb_l\bb_l^*\BZ$ on $\OMB$ equals to $0$. Therefore each $z_l$ is a bounded and centered random variable. We can use scalar version of  Bernstein inequality (a special version of Theorem \ref{BernFourier}) to estimate $\sum_{l\in\Gamma_p} z_l.$
First we estimate $|z_l|,$
\begin{equation*}
| z_l | \leq  | \be_s^* \bphi_l \bphi_l^* \bz | = |\be_s^* \bphi_l| \cdot |\bphi_l^* \bz| \leq \frac{\mu_{\max}}{\sqrt{L}} |\bphi_l^*  \bz|.
\end{equation*}

There holds $|\be_s^*\bphi_l| \leq  \|\bba_l\otimes \bb_l \|_{\infty} = \|\bba_l\|_{\infty} \|\bb_l\|_{\infty}  \leq \frac{\mu_{\max}}{\sqrt{L}}$, which follows from the fact that $\ba_l$ is a row of DFT matrix and each entry of $\ba_l$ is of unit magnitude. Moreover, $\bz$ has support $\OM$,
\begin{equation*}
|z_l| \leq \frac{\mu_{\max}}{\sqrt{L}} |\bphi_l^*  \bz| 
= \frac{\mu_{\max}}{\sqrt{L}}|\bphi_{l,\OM}^*  \bz| 
\leq \frac{\mu_{\max}}{\sqrt{L}}\| \bphi_{l,\OM}\|\cdot \| \bz\| 
\leq  \frac{\mu^2_{\max}}{\sqrt{L}} \cdot \sqrt{\frac{kn}{L}} \|\bz\| 
\leq \frac{\mu^2_{\max}\sqrt{kn}\|\bz\|}{L}.
\end{equation*} 
Therefore, $R: = \max_{l}|z_l| \leq \frac{\mu^2_{\max}\sqrt{kn}\|\bz\|}{L}.$ The variance of $\sum_{l\in\Gamma_p}z_l$ is estimated as 
\begin{eqnarray*}
\E |z_l|^2 & = & \E|  \be_s^* \bphi_l\bphi_l^*  \bz  |^2  =  \E | \be_s^* \bphi_l |^2 |\bphi_l^*  \bz|^2 \\
& \leq & \frac{\mu^2_{\max}}{L} \E|\bphi_l^* \bz|^2 = \frac{\mu^2_{\max}}{L} \bz^* ( \I_N \otimes \bb_l\bb_l^* ) \bz
\end{eqnarray*}
and
\begin{equation*}
\sigma^2 = \sum_{l\in\Gamma_p} \E | z_l |^2 \leq \frac{\mu^2_{\max}}{L}\bz^* (\I_N \otimes \BT_p) \bz \leq
\frac{5\mu^2_{\max}Q\|\bz\|^2}{4L^2}.
\end{equation*}

Now we apply the Bernstein inequality in Theorem \ref{BernFourier} and have the estimate for $\sum_{l\in\Gamma_p} z_l,$
\begin{equation*}
\Pr\left( | \sum_{l\in\Gamma_p} z_l |> t \right) \leq 2\exp \left( -\frac{t^2/2}{ 5\mu^2_{\max}Q\|\bz\|^2/4L^2 +
\mu^2_{\max}\sqrt{kn} \|\bz\| t/ 3L} \right).
\end{equation*}
Note that   $\BW_{p-1}$ is independent of $\A_p^*\A_p$ and hence the inequality above is still valid if we choose $\bz = \VEC(\BW_{p-1})$.  Since $\|\bz\| \leq 2^{-p+1} \sqrt{kn}$ and $t = 2^{-p-1} \frac{Q}{L}$, we have
\begin{equation*}
\Pr\left( | \sum_{l\in\Gamma_p} z_l |> \frac{Q}{2^{p+1}L} \right) \leq 2\exp \left( -\frac{3Q}{128\mu^2_{\max}kn}
\right).
\end{equation*}
We take the union bound over all $s\in \OMB$ and obtain
\begin{equation*}
\Pr\left( \| \PP_{\OMB}\A_p^*\A_p(\BW_{p-1})\|_{\infty} > \frac{Q}{2^{p+1}L} \right) \leq 2kN \exp \left(
-\frac{3Q}{128\mu^2_{\max}kn} \right).
\end{equation*}
\end{proof}

\section{Proof of Theorem~\ref{main_thm_noise}}\label{s:stability}

Now we consider the noisy case by solving the convex program \eqref{l1_noise_matrix}.  Our goal is to prove Theorem~\ref{main_thm_noise}.
The main ingredient of our proof is Theorem 4.33 in \cite{FouRa13}. A variant of that theorem, adapted to our
setting, reads as follows.

\begin{Theorem}\label{433}
Let $\tbphi_{s,t}$ be the columns of $\BPhi\in \CC^{L\times kN}$, $\bv_0 = \VEC(\BX_0)\in \CC^{kN\times 1}$ with
support on $\OM$ and $\by = \BPhi \bv_0 + \bw$ with $\|\bw\|\leq \eta$. For $\delta,\beta,\theta,\eta,\tau \geq 0$ and $\delta < 1$,
\begin{equation}
\|\BPhi_{\OM}^* \BPhi_{\OM} - \I_{\OM}\| \leq \delta, \quad \max_{1\leq s\leq k, t> n} \| \BPhi_{\OM}^* \tbphi_{s,t} \| \leq \beta,
\end{equation}
and that there exists a vector $\bq = \BPhi^* \bp \in \CC^{kN\times 1}$ with $\bp\in \CC^L$ such that
\begin{equation}
\| \bq_{\OM} - \sgn(\bv_0) \| \leq \frac{1}{4\sqrt{2}\gamma}, \quad \| \bq_{\OMB} \|_{\infty} \leq \theta, \quad \| \bp \| \leq \tau \sqrt{kn}.
\end{equation}
If $\rho: \theta + \frac{\beta}{4\sqrt{2}\gamma (1  - \delta)} < 1$, then the minimizer $\hbv$ to \eqref{l1_noise_matrix} satisfies 
\begin{equation}
\| \hbv - \bv_0 \| \leq (C_0 + C_1 \tau \sqrt{kn} )\eta
\end{equation}
where $C_0$ and $C_1$ are two scalars only depending on $\delta, \beta, \gamma, \theta$ and $\tau.$
\end{Theorem}

Actually, we have already proven in the previous section that we can choose $\delta = \frac{1}{2},  \theta =
\frac{1}{2}$.
 The only two unknown components are  $\tau$ and $\beta$. $\tau$ follows from computing the norm of $\bp.$ We need to know the coherence $\mu$ of $\BPhi$ to estimate $\beta$ and claim that $\beta \leq 1$.
Define the coherence as the largest correlation between two different columns of the $L \times kN$ matrix $\BPhi$,
\begin{equation}
\mu := \max_{ i_1\neq i_2 \text{ or } j_1 \neq j_2} | \lag \tbphi_{i_1,j_1}, \tbphi_{i_2, j_2} \rag|, \quad \tbphi_{s,t} = \diag(\tbb_s) \tba_t
\end{equation}
where $\tbphi_{s,t}$ is defined in~\eqref{def:phist}.

\subsection{Estimation of $\mu$ and $\beta$.}

It is easy to obtain a bound of $\|\BPhiO^*\tbphi_{s,t}\|$ for $1\leq s\leq k$ and $t > n$ with  $\mu$, 
\begin{equation*}
\|\BPhiO^*\tbphi_{s,t}\| \leq \sqrt{kn} \mu.
\end{equation*}

It suffices to bound $\mu$ by $1/\sqrt{kn}$ in order to achieve $\beta \leq  1$.
The following lemma (see Lemma A.9 in \cite{FouRa13}) will be used in the proof.
 
\begin{lemma}\label{coherencebound}
For any matrix $\BC$, any entry is bounded by its operator norm, i.e.,
\begin{equation}
\max_{i,j}|\BC_{ij}| \leq \|\BC\|.
\end{equation}
\end{lemma}

\begin{Theorem}\label{beta}
There holds
\begin{equation*}
\max_{t>n}\|\BPhiO^*\tbphi_{s,t}\| \leq 1
\end{equation*}
with probability at least $1 - L^{-\alpha}$ if 
\begin{enumerate}
\item $\BA$ is a Gaussian random matrix and  $L\geq 4C_{\tilde{\alpha}} \mu^2_{\max}\max\{kn \log(L), \sqrt{kn}\log^2L \}$ and $\tilde{\alpha} = 2\log(kN) + \alpha $.
\item $\BA$ is a random Fourier matrix and  $L\geq 4C_{\tilde{\alpha}} \mu^2_{\max}kn \log(L)$ and $\tilde{\alpha} = 2\log(kN) + \alpha $.
\end{enumerate}
Here, in both cases $C_{\tilde{\alpha}}$ is a function growing at most linearly with respect to $\tilde{\alpha}$.
\end{Theorem}

\begin{proof}
We only prove the Gaussian case by using Lemma \ref{weakrip}. The proof of the Fourier case is exactly the same. First we pick any two different columns $\tbphi_{i_1,j_1}$ and $\tbphi_{i_2,j_2}$ from $\BPhi$, where $1\leq i_1, i_2\leq k$ and $1\leq j_1, j_2\leq N$.   Let $\tOM$ be
\begin{equation}
\tOM = \{ (i_1,j_1), (i_1, j_2), (i_2, j_1), (i_2,j_2) \}.
\end{equation}
 $|\tOM|\leq 4$ and $\BPhi_{\tOM}$ contains  $\tbphi_{i_1,j_1}$ and $\tbphi_{i_2,j_2}$. By Lemma \ref{weakrip},
$\|\BPhi^*_{\tOM} \BPhi_{\tOM} - \I_{\tOM}\| \leq \delta$ with probability $1 - L^{-\talpha}$ if $L \geq
4C_{\talpha} \max\{ \log L/\delta^2, \log^2 L/\delta \} $. Therefore $|\lag \tbphi_{i_1,j_1},
\tbphi_{i_2,j_2}\rag| \leq \delta$ follows from Lemma \ref{coherencebound}. Then we take the union bound over $(kN)^2/2$ pairs of indices $(i_1, j_1)$ and $(i_2,j_2)$.
Therefore, $ \mu = \max_{i_1,i_2,j_1.j_2}| \lag \tbphi_{i_1,j_1}, \tbphi_{i_2,j_2} \rag | \leq 1/\sqrt{kn}$ 
 with $L \geq 4C_{\talpha} \max\{kn \log L, \sqrt{kn}\log^2L\}$ and probability at least $1 - (kN)^2L^{-\talpha} \geq 1 - L^{-\alpha}$ if $\talpha = 2\log (kN) + \alpha$.
\end{proof}

\subsection{Proof of Theorem \ref{main_thm_noise} when $\BA$ is a Gaussian random matrix}

\begin{proof}
Let $\bp = \BPhiO(\BPhiO^*\BPhiO)^{-1}\sgn(\bv_0)$ and $\bq = \BPhi^*\bp$ as defined in \eqref{gaussian_dual}. From
 Lemma \ref{weakrip}, Theorem \ref{gaussian_dual_inf_norm} and  Theorem~\ref{beta} we have
\begin{equation*}
\| \BPhi_{\OM}^* \BPhi_{\OM} - \I_{\OM} \| \leq \delta=  \frac{1}{2}, \quad \theta = \|\bq_{\OMB}\|_{\infty} < \frac{1}{2}, \quad \beta \leq 1
\end{equation*}
with probability at least $1 - O(L^{-\alpha + 1})$.
\begin{equation*}
\|\bp\|^2 = \sgn(\bv_0)^T ( \BPhi_{\OM}^* \BPhi_{\OM} )^{-1} \sgn(\bv_0) \leq 2kn \Longrightarrow \|\bp\| \leq \sqrt{2kn}, \quad \tau = \sqrt{2}
\end{equation*}
and also
\begin{equation*}
\rho = \theta + \frac{\beta}{4\sqrt{2}\gamma (1 - \delta)} \leq \frac{1}{2} + \frac{1}{2\sqrt{2}\gamma} < 1,
\end{equation*}
where $\gamma$ is the operator norm of $\A$ and is certainly greater than $1$. Applying Theorem \ref{433} gives the desired result. 
\end{proof}

\subsection{Proof of Theorem \ref{main_thm_noise} when $\BA$ is a random Fourier matrix}

\begin{proof}
We have an explicit form of $\bp$ from \eqref{dual_cert_w}  \begin{equation*}
\bp = -\frac{L}{Q}( \A_1(\BW_0), \cdots, \A_P(\BW_{P-1}) )^T, 
\end{equation*}
such that $\bq = \VEC(\BY)$, $\BY = \A^*(\bp) = -\frac{L}{Q} \sum_{p=1}^P \A_p^*\A_p(\BW_{p-1})$ and $\BY$ is an inexact dual certificate. 
The norm of $\bp$ is
\begin{equation*}
\|\bp\|^2 = \frac{L^2}{Q^2} \sum_{i=1}^P \|\A_p(\BW_{p-1})\|_F^2.
\end{equation*}
For each $\A_p(\BW_{p-1})$,
\begin{equation*}
\| \A_p(\BW_{p-1}) \|^2 = \lag \A_{p,\OM}^*\A_{p,\OM} \BW_{p-1}, \BW_{p-1} \rag \leq \frac{3Q}{2L} \|\BW_{p-1}\|_F^2 < \frac{2Q}{L} 4^{-p+1} kn,
\end{equation*}
which follows from \eqref{wnorm}  and $\| \A_{p,\OM}^*\A_{p,\OM}\| \leq \frac{3Q}{2L}$ in Lemma \ref{lowbound}.
Therefore,
\begin{equation*}
\|\bp\| \leq \sqrt{\frac{2L}{Q}} \sum_{p=1}^P 2^{-p+1} \sqrt{kn} \leq 2\sqrt{2P} \cdot \sqrt{kn}.
\end{equation*}
This implies  $\tau = 2\sqrt{2P}$ where $P$ is chosen in \eqref{PP}.
We also have $\delta = \frac{1}{2}$, $\theta = \frac{1}{2}$ and $\beta \leq 1$ which are the same as Gaussian case and $\rho = \theta + \frac{\beta}{4\sqrt{2}\gamma(1 - \delta)} < 1.$
Now we have everything needed to apply Theorem \ref{433}. 	
The solution $\hat{\BX}$ to \eqref{l1_noise_matrix} satisfies
\begin{equation*}
 \|\hat{\bv} - \bv_0 \|_F = \|\hat{\BX} - \BX_0 \|_F  \leq (C_0 + C_1 \sqrt{P} \sqrt{kn} )\eta,
\end{equation*}
with $P = O(\log(N))$, and $C_0$, $C_1$ are two constants. 
\end{proof}

\section{Numerical simulations}
\label{s:numerics}

\subsection{Recovery performance of SparseLift }

\begin{figure}[h]
\begin{minipage}{0.48\textwidth}
\centering
\includegraphics[width = 90mm]{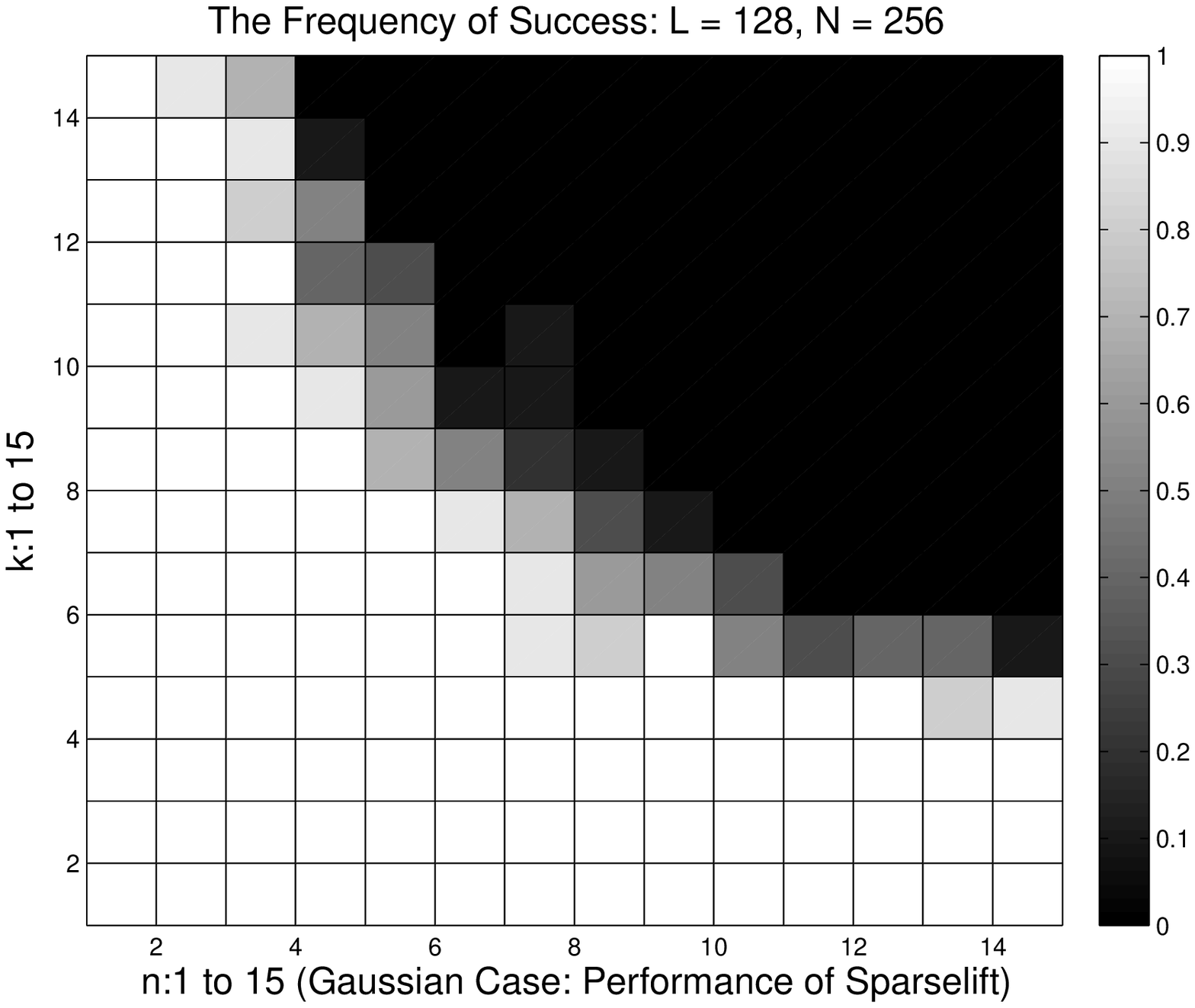}
\end{minipage}
\hfill
\begin{minipage}{0.48\textwidth}
\centering
\includegraphics[width = 90mm]{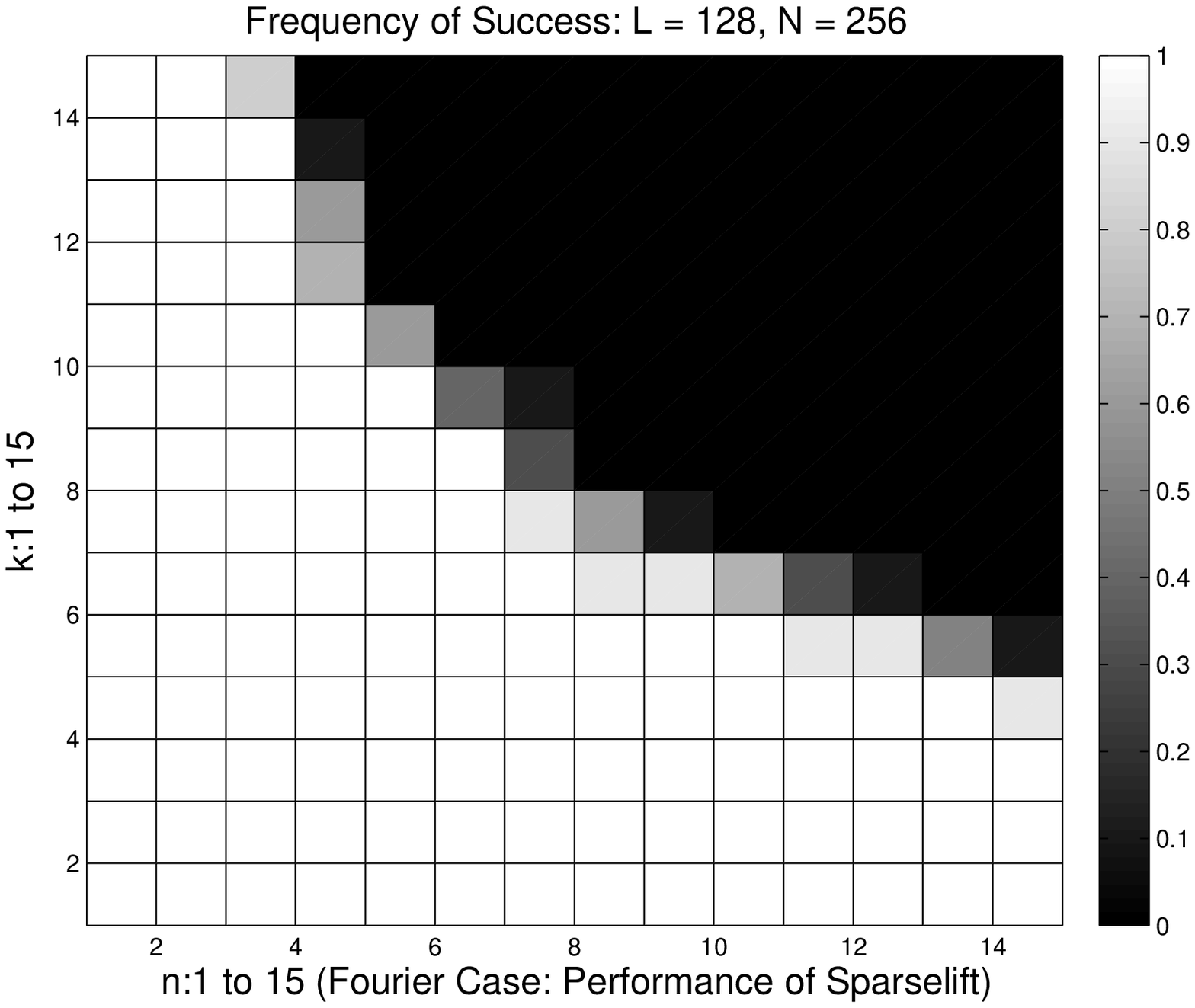}
\end{minipage}
\caption{Phase transition plot of performance by solving SparseLift \eqref{l1} directly. }
\label{fig:l1_recovery}
\end{figure}

We investigate the empirical probability of success of SparseLift over different pairs of $(k,n)$ for fixed $L$ and $N$. We  fix $L = 128$, $N = 256$ and make both $k$ and $n$ range  from $1$ to $15$.  We choose $\BB$ as the first $k$ columns of an $L\times L$ unitary DFT matrix and $\BA$ is sampled independently for each experiment. $\bx$ has random supports with cardinality $n$ and the nonzero entries of $\bh$ and $\bx$ yield $\mathcal{N}(0,1)$. For each $(k,n)$, we run $\eqref{l1}$ 10 times and compute the probability of success. We classify a recovery as a success if the relative error is less than $1\%$,
\begin{equation*}
\frac{\| \hBX - \BX_0 \|_F}{\|\BX_0\|_F} \leq 0.01, \quad \BX_0 = \bh_0\bx_0^T.
\end{equation*}
Note that SparseLift is implemented by using CVX package~\cite{cvx}.  The results are depicted in Figure \ref{fig:l1_recovery}.
The phase transition plots are similar for  the case when $\BA$ is a Gaussian random matrix or a random Fourier matrix.
In this example choosing $kn < 70$ gives satisfactory recovery performance via solving SparseLift. 

\begin{figure}[h]
\begin{minipage}{0.48\textwidth}
\centering
\includegraphics[width = 90mm]{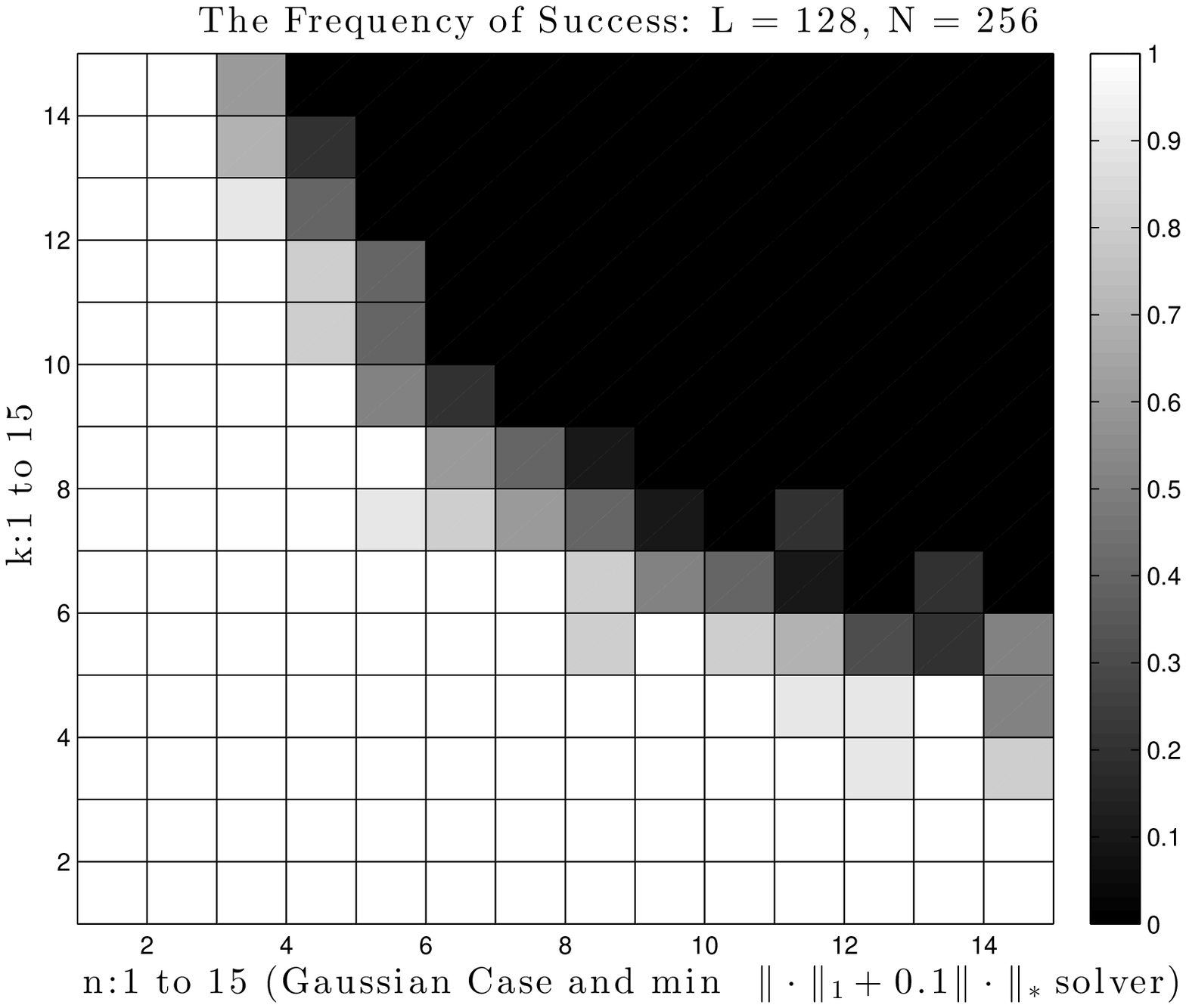}
\end{minipage}
\hfill
\begin{minipage}{0.48\textwidth}
\centering
\includegraphics[width = 90mm]{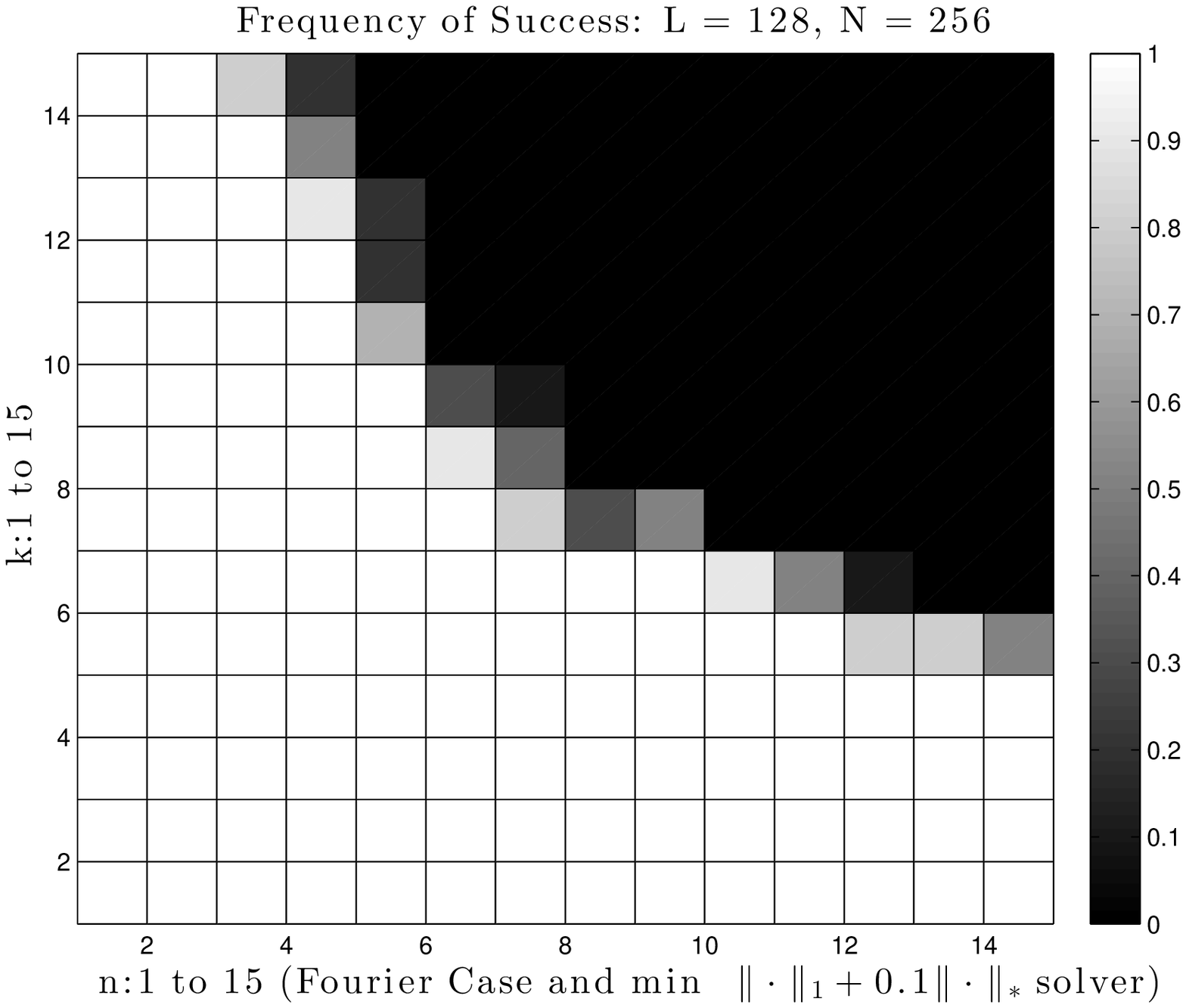}
\end{minipage}
\caption{Phase transition plot of performance via $\|\cdot\|_1 + \lambda \|\cdot\|_*$ minimization for
$\lambda = 0.1$.}
\label{l1+0dot1nuc}
\end{figure}

\begin{figure}[h]
\begin{minipage}{0.48\textwidth}
\centering
\includegraphics[width = 90mm]{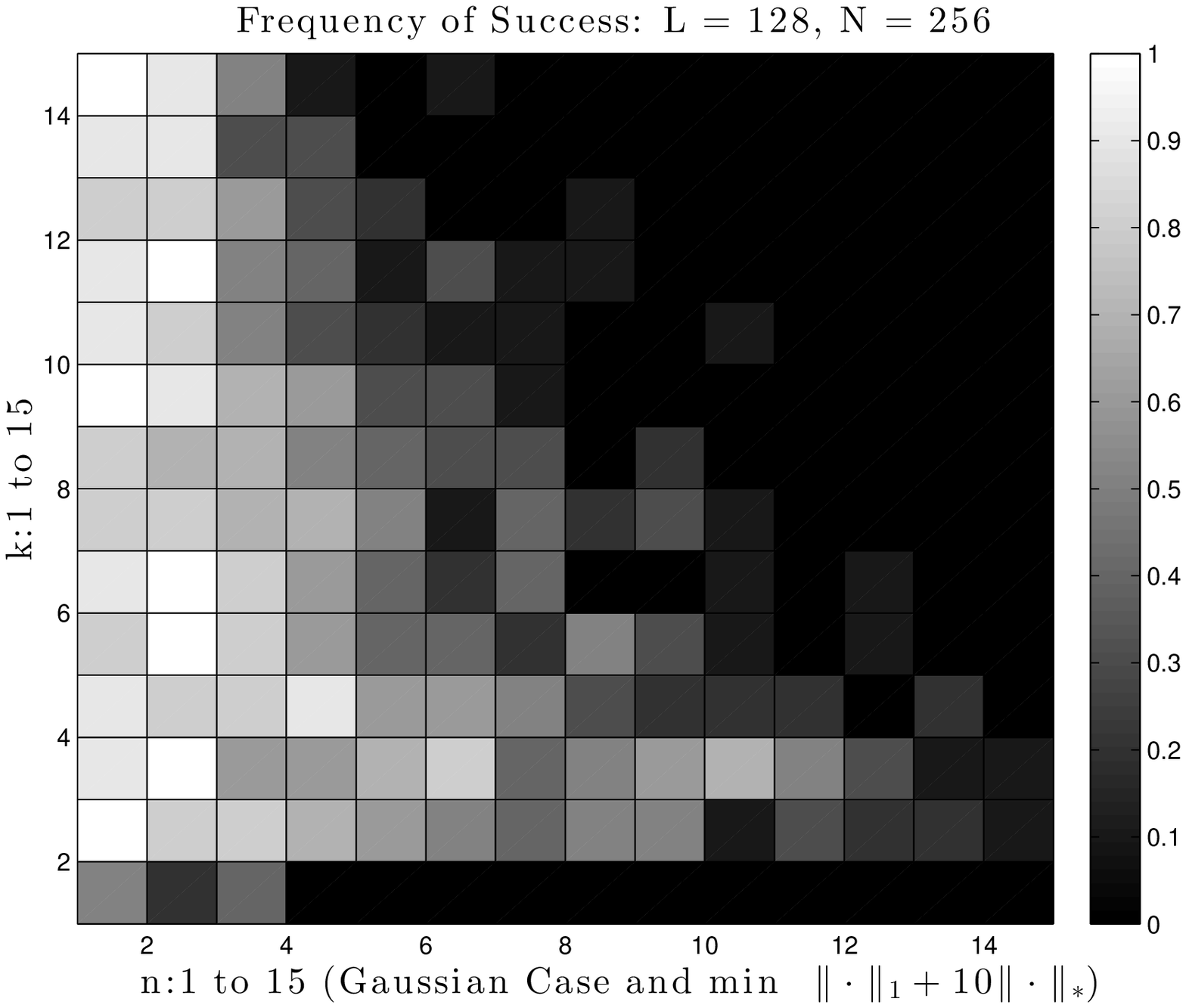}
\end{minipage}
\hfill
\begin{minipage}{0.48\textwidth}
\centering
\includegraphics[width = 90mm]{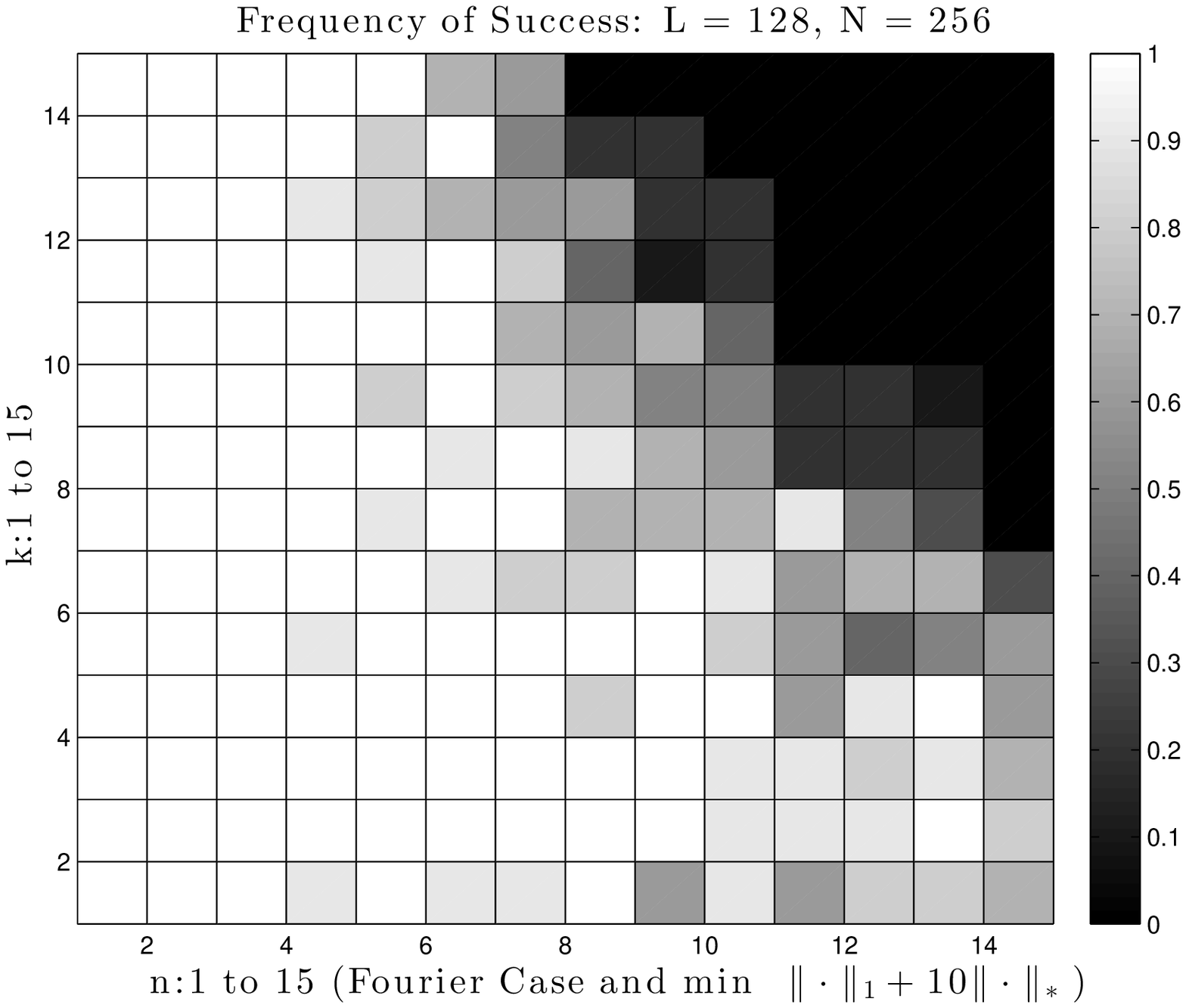}
\end{minipage}
\caption{Phase transition plot of performance via $\|\cdot\|_1 + \lambda \|\cdot\|_*$ minimization for
$\lambda = 10$.}
\label{l1+10nuc}
\end{figure}

For comparison, we illustrate the performance when we solve the convex optimization problem that combines the nuclear norm and sparsity as in~\eqref{l1+lambdanuc}, see Figures \ref{l1+0dot1nuc} and \ref{l1+10nuc}. The values for $\lambda$ are $\lambda = 0.1$ (the best value for $\lambda$ for this experiment) and $\lambda = 10$. The settings of other parameters are exactly the same as the scenario of solving SparseLift. Figure \ref{l1+0dot1nuc} shows that \eqref{l1+lambdanuc} with $\lambda = 0.1$ performs similar to
 solving $\eqref{l1}$. However, if $\lambda=10$ the results are clearly worse in general, in particular for the Gaussian case, see Figure \ref{l1+10nuc}. 
For Fourier case, choosing $\lambda = 10$ might do better than SparseLift for some pairs of $(k,n)$. The reason behind this phenomenon is unclear yet and may be due to solver we choose and numerical imprecision. 
All the numerical results support our hypothesis that  the performance of recovery is sensitive to the
choice of $\lambda$ and SparseLift is good enough to recover calibration parameters and the signal in most cases with theoretic guarantee. This is
in line with the findings in~\cite{OJF12} which imply that convex optimization with respect to a combination of
the norms $\|\cdot\|_1$ and $\|\cdot\|_*$ will not perform significantly better than using only one norm,
the $\ell_1$-norm in this case.

We also show the performance when using a mixed $\ell_2$-$\ell_1$-norm to enforce column-sparsity in $\BX$
instead of just $\ell_1$-norm. In this case we note a slight improvement over~\eqref{l1}.  As mentioned earlier, analogous versions of our main results,
Theorem~\ref{th:main} and Theorem~\ref{main_thm_noise}, hold for the mixed $\ell_2$-$\ell_1$-norm case as well and will be reported in a forthcoming paper.

\begin{figure}[h]
\begin{minipage}{0.48\textwidth}
\centering
\includegraphics[width = 90mm]{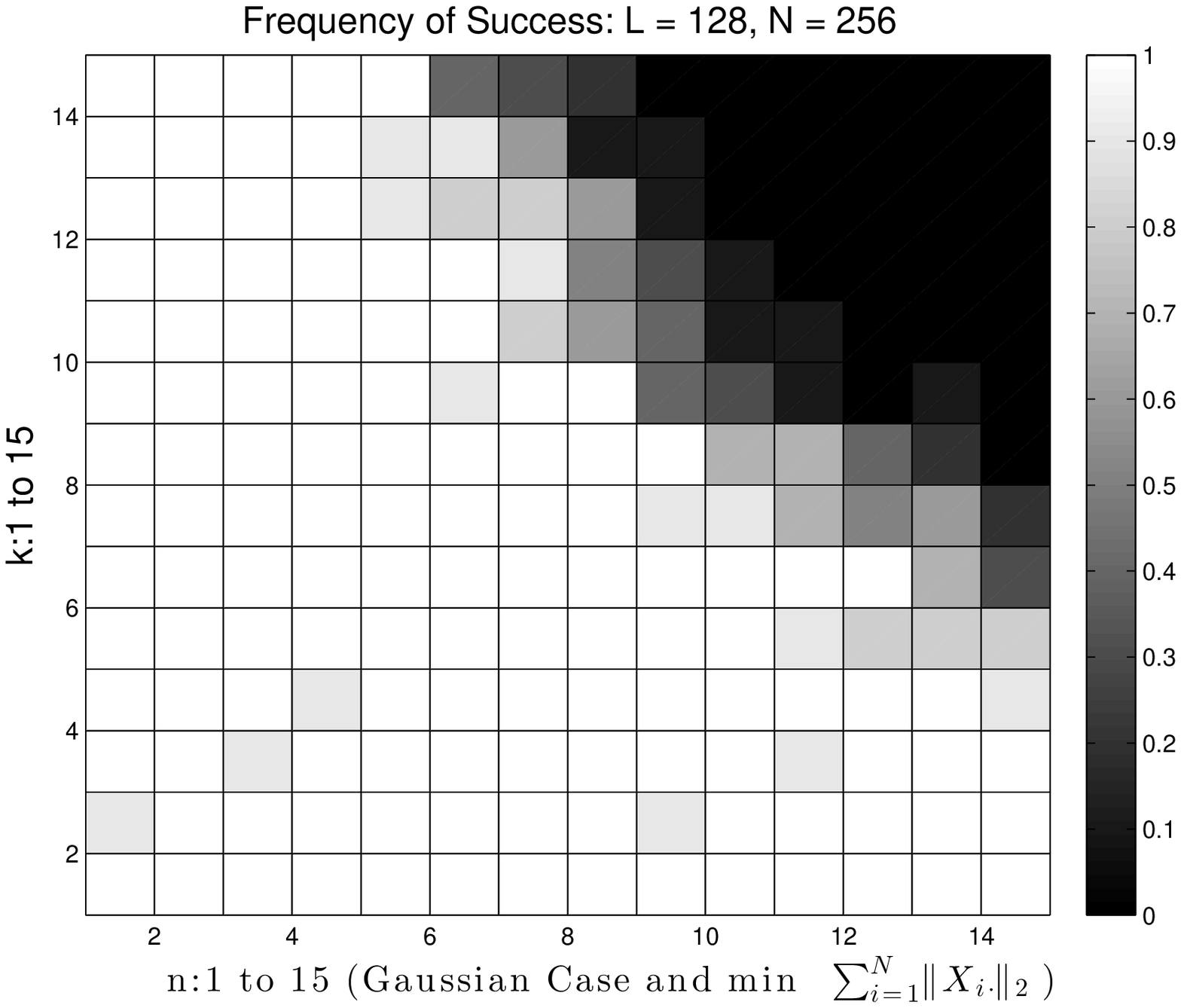}
\end{minipage}
\hfill
\begin{minipage}{0.48\textwidth}
\centering
\includegraphics[width = 90mm]{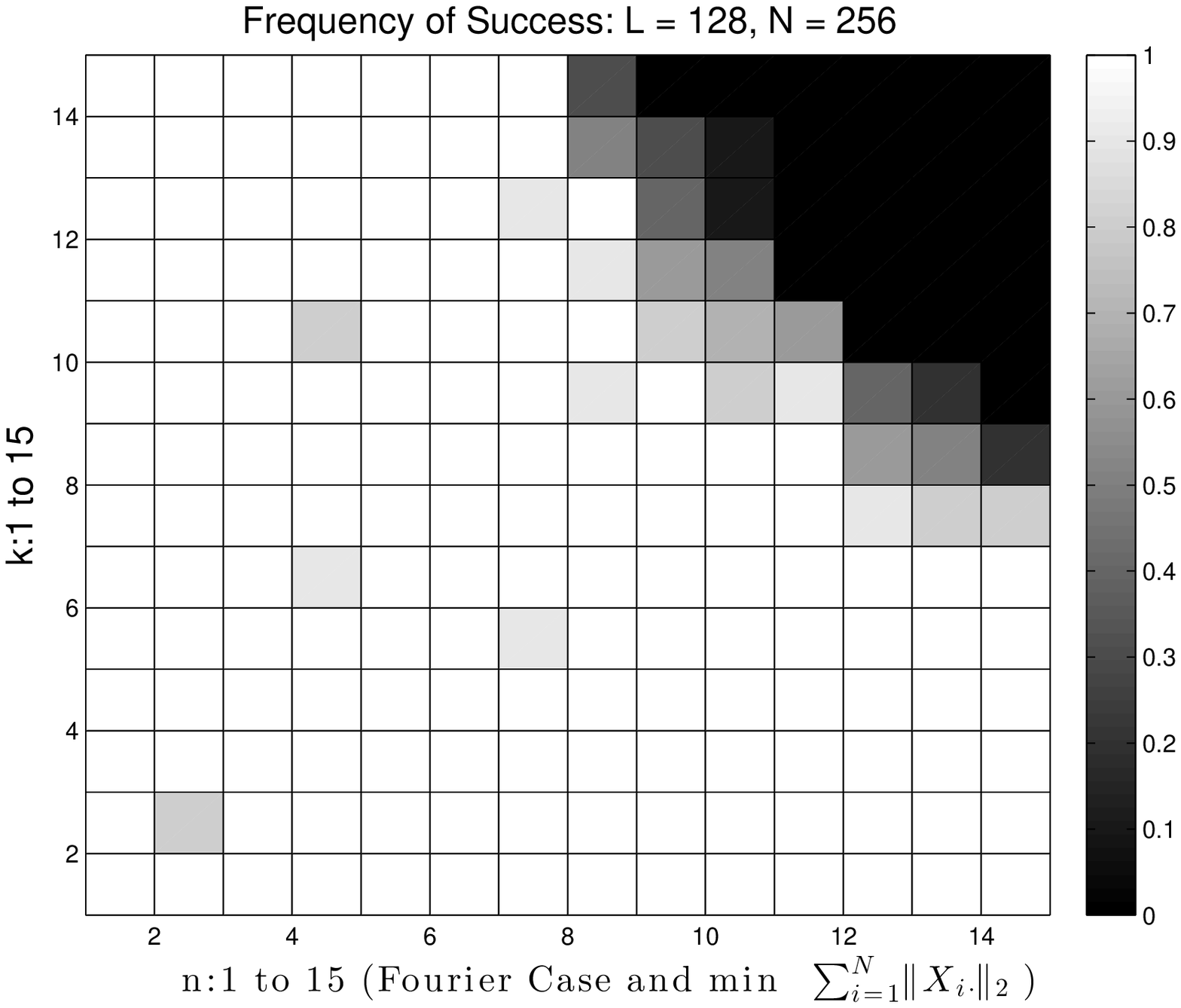}
\end{minipage}
\caption{Phase transition plot of performance via  $\|\cdot\|_{1,2}$ minimization. }
\label{l1+0dot01nuc}
\end{figure}

\subsection{Minimal $L$ required for exact recovery is  proportional to $kn$}
\begin{figure}[h]
\begin{minipage}{0.48\textwidth}
\centering
\includegraphics[width = 90mm]{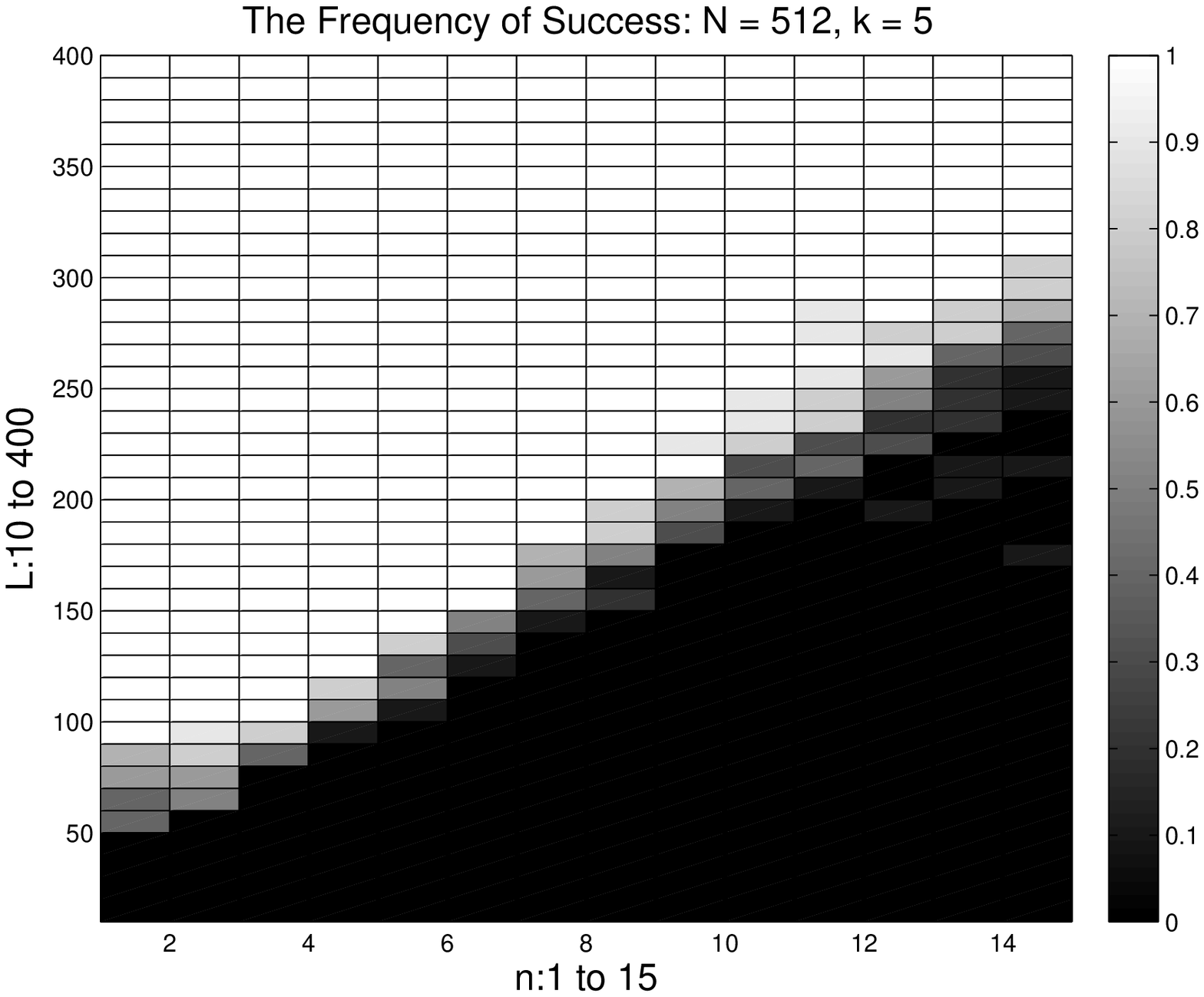}
\end{minipage}
\hfill
\begin{minipage}{0.48\textwidth}
\centering
\includegraphics[width = 90mm]{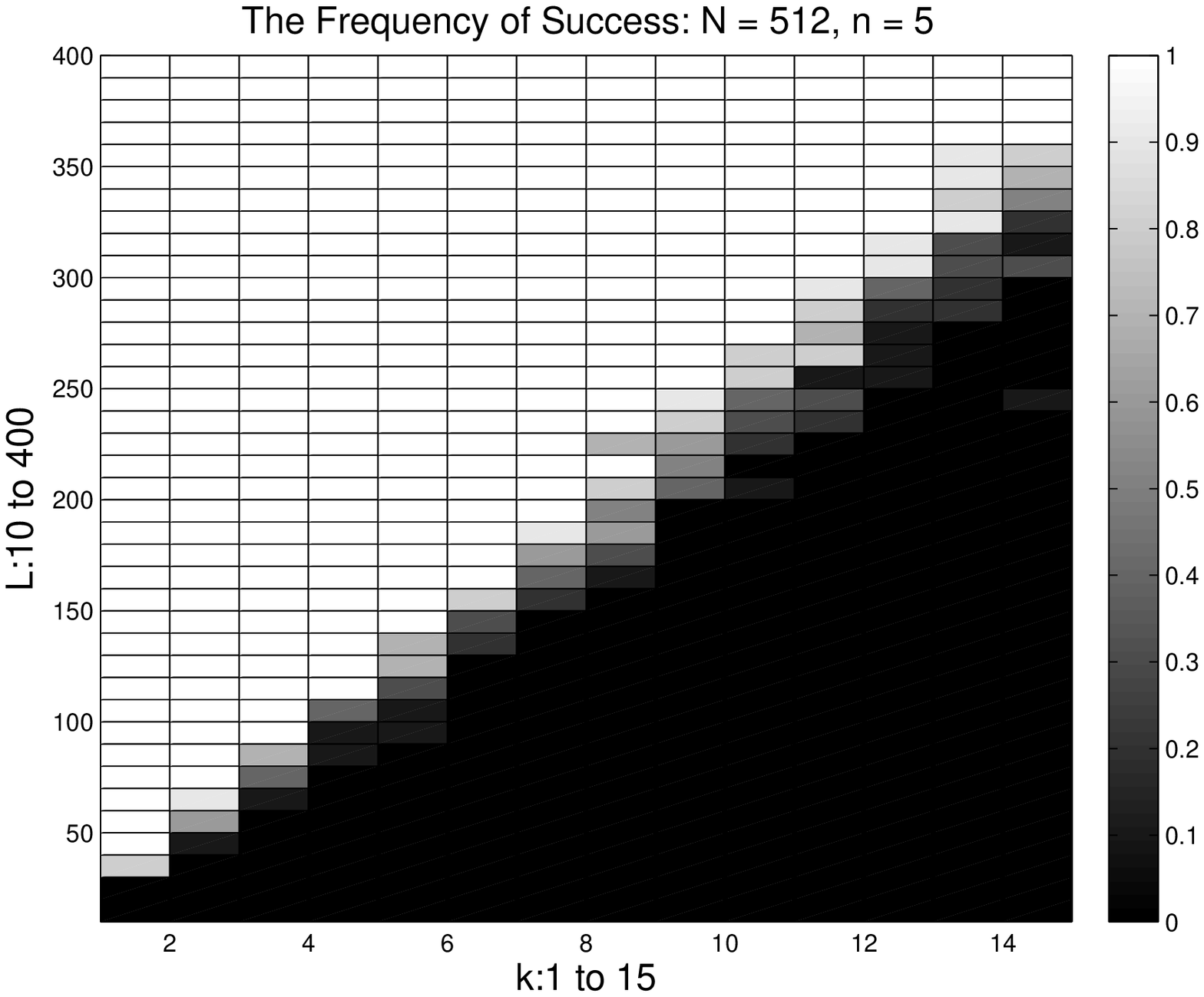}
\end{minipage}
\caption{Phase transition plot of performance via  SparseLift (implemented by using Yall1~\cite{yall1}). Here $N = 512$ and $L$ varies from 10 to 400 for each pair of $(k,n)$. The left figure: $k = 5$ and $n$ is from 1 to 15; the right figure: $n=5$ and $k$ is from 1 to 15. }
\label{l1_fixed_kn}
\end{figure}

The minimal $L$ which guarantees exact recovery is nearly proportional to $kn$, as shown in Figure~\ref{l1_fixed_kn}. The figure shows the results of two experiments: one is to keep $k=5$ and let $n$ vary from 1 to 15; the other is to fixe $n=5$ and let $k$ range from 1 to 15. In both experiments, $N$ is chosen to be $512$, $L$ varies from 10 to 400 and $\BA$ is chosen to be a Gaussian random matrix. We run 10 simulations  for each set of $(k,n,L)$ and treat a recovery as a success if the relative error is less than $1\%.$ The result provides numerical evidence for the theoretical guarantee that SparseLift yields exact recovery if $L \geq C_{\alpha} kn \log^2 L\log (kN)$. 

We also notice that the ratio of minimal $L$ against $kn$ is slightly larger in the case when $n$ is fixed and $k$ varies than in the other case. This difference can  be explained by  the $\log(kN)$-term because the larger $k$ is, the larger $L$ is needed.

However, these numerical observations do not imply that the minimum $L$ required by a different (numerically feasible) algorithm also has to be
proportional to $kn$. Indeed, it would be very desirable to have an efficient algorithm with provable recovery guarantees for which on the order of
about $k+n$ (modulo some log-factors) measurements suffice.

\subsection{Self-calibration and direction-of-arrival estimation}
\label{ss:arraynumerics}

We examine the array self-calibration problem discussed in Section~\ref{ss:doa}.
We consider a scenario of three fully correlated equal-power signals. Note that the case of fully correlated
signals
is {\em much harder} than the case of uncorrelated (or weakly correlated) signals. 
Standard methods like MUSIC~\cite{WF89} will in general fail completely for
fully correlated signals even in the absence of any calibration error. 
The reason is that these methods rely on
the additional information obtained by taking multiple snapshots. Yet, for fully correlated signals all
snapshots are essentially identical (neglecting noise) and therefore no new information is contained in 
multiple snapshots.

We use a circular array consisting of $L=64$ omni-directional sensors.
We assume that the calibration error is changing slowly across sensors, as for example in the case of
sensors affected by changing temperature or humidity. There are of course many possibilities
to model such a smooth change in calibration across sensors. In this simulation we assume that the variation can be
modeled by a low-degree trigonometric polynomial (the periodicity inherent in trigonometric polynomials
is fitting in this case, due to the geometry of the sensors); in this example we have chosen a polynomial of degree 3.
To be precise, the calibration error, i.e., the complex gain vector $\bd$ is given by $\bd=\BB\bh$, where $\BB$ is 
a $64 \times 4$ matrix, whose columns are the first four columns of the $64 \times 64$ DFT matrix and
$\bh$ is a vector of length 4, whose entries are determined by taking a sample of a multivariate complex Gaussian 
random variable with zero mean and unit covariance matrix.
The stepsize for the discretization of the grid of angles for the matrix $\BA$ is 1 degree, hence $N = 180$ (since we consider angles between
$-90^{\circ} $ and $90^{\circ}$).
The directions of arrival are set to $-10^{\circ}, 5^{\circ}$,
and $20^{\circ}$, and the SNR is 25dB. Since the signals are fully correlated we take only one snapshot.

\begin{figure}[h]
\centering
\includegraphics[width = 90mm]{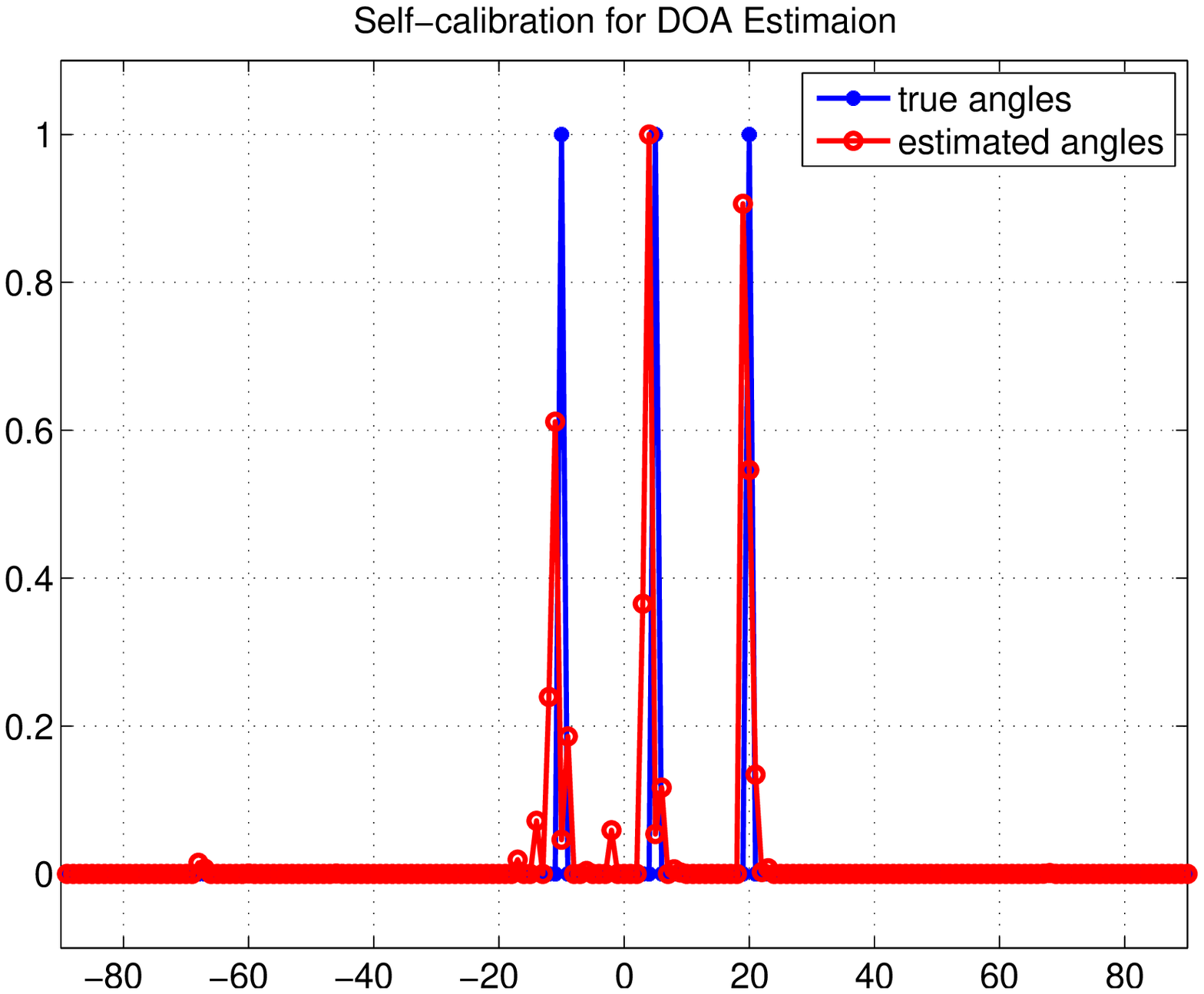}
\caption{Estimated angles of arrival using self-calibration via SparseLift.}
\label{fig:DOA}
\end{figure}

We apply SparseLift, cf.~\eqref{l1_noise_matrix}, to recover the angles of arrival. Since the sensing matrix $\BA$ does not fulfill the conditions 
required by Theorem~\ref{main_thm_noise} and  the measurements are noisy, we cannot expect SparseLift to return
a rank-one matrix. We therefore extract the vector with the angles of arrival by computing the right singular
vector associated with the largest singular value of $\hat{\BX}$. 
As can be seen in Figure~\ref{fig:DOA} the estimate obtained via self-calibration provides a quite accurate 
approximation to the true solution.

\subsection{5G, the Internet of Things, and SparseLift} 
\label{ss:5Gnumerics}

We briefly discussed in Section~\ref{ss:5G} how self-calibration problems of the form analyzed in this paper
are anticipated to play a role in 5G wireless communication systems in connection with the Internet of Things,
sporadic traffic and random access. Let us assume we deal with a CDMA communication scheme. To accommodate as many sporadic users 
as possible, we consider an ``overloaded'' CDMA system. This means that the matrix $\BA$ containing the so-called
spreading sequences (aptly called spreading matrix) has many more columns than rows. 
We use a random Fourier matrix for that purpose.
Furthermore, we assume that the transmission channel is frequency-selective and time-invariant for the time
of transmission of one sequence, but it may change from transmission to transmission. The channel is unknown
at the receiver, it only knows an upper bound on the delay spread (the length of the interval containing the
non-zero entries of $\bh$). Moreover, the transmitted
signal experiences additive Gaussian noise. The mathematician who is not familiar with all this 
communications jargon may rest assured that this is a standard scenario in wireless communication. 
In essence it translates to a linear system of equations of the form $\by = \BD \BA \bx + \bw$,
where $\BA$ is a flat random Fourier matrix, $\BD$ is of the form $\BD = \diag(\BB \bh)$ and $\bw$
is AWGN. To be specific, we choose the transmitted signal $\bx$ to be of length $N=256$ with $n=5$ 
data-bearing coefficients (thus, $\|\bx\|_0 = 5$), the locations of which are chosen at random. 
The 256 spreading sequences of length $L=128$  are generated by randomly selecting
128 rows from the $256\times 256$ DFT matrix. Thus the system is {\em overloaded} by a factor of 2.
The delay spread is assumed to be 5, i.e., $k=5$ and the $L  \times k$ matrix $B$ consists of the first 5 columns of the unitary $L \times L$ DFT matrix.
The SNR ranges from 0dB to 80dB (the latter SNR level is admittedly unrealistic in practical scenarios).

We attempt to recover $\bx$ by solving \eqref{l1_noise}.  
Here we assume $\bw_i \sim \mathcal{N}(0, \sigma^2)$ and $\|\bw\|^2/\sigma^2 \sim \chi^2_{L}.$ 
We choose $\eta = (L + \sqrt{4L})^{1/2}\sigma$ as \cite{RechtRom12} in order to make $\BX_0$ 
inside the feasible set. 

\begin{figure}[h]
\begin{minipage}{0.48\textwidth}
\centering
\includegraphics[width = 90mm]{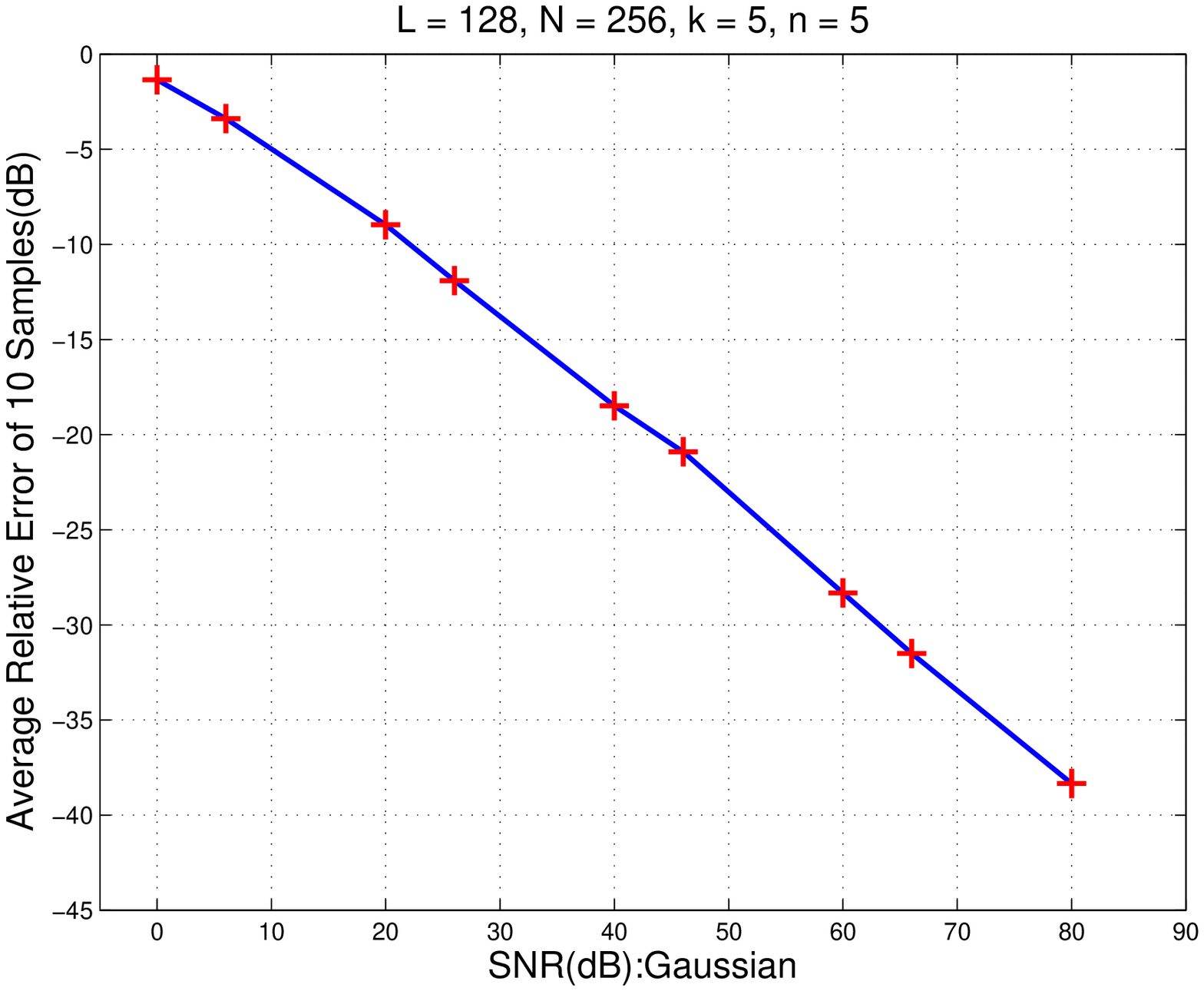}
\end{minipage}
\hfill
\begin{minipage}{0.48\textwidth}
\centering
\includegraphics[width = 90mm]{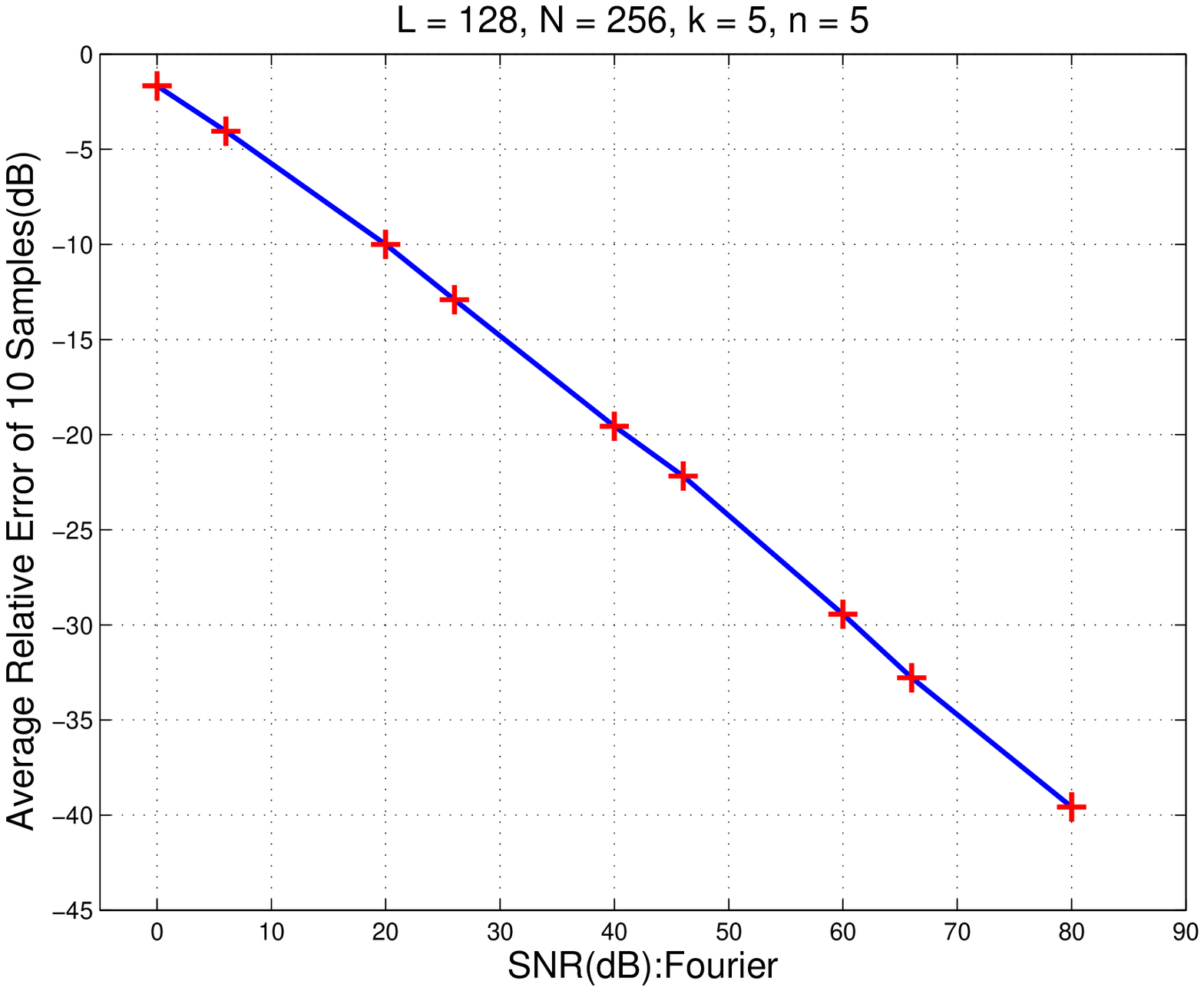}
\end{minipage}
\caption{Sporadic traffic in 5G wireless communications: Performance of SparseLift in the  presence of noise for
different types of $\BA$. }
\label{SNR}
\end{figure}

The result is depicted in Figure~\ref{SNR}; for comparison we also show the results when using a Gaussian random
matrix as spreading matrix.  We plot the relative error (on a dB scale) for different levels of $\SNR$ (also
on a dB scale).  The horizontal axis represents $\SNR = 10\log_{10} \left( \frac{\|\BX_0\|_F^2}{\|\bw\|^2}
\right)$ and the vertical axis represents $10\log_{10} \left( \text{Avg.Err.}\right)$. We run 10 independent 
 experiments for each SNR level and take the average relative error.  The
error curves show clearly the desirable linear behavior between SNR
and mean square error with respect to the log-log scale.

\section{Conclusion and outlook}
\label{s:conclusion}

Self-calibration is an emerging concept that will get increasingly important as hardware and software will become more and more intertwined. 
Several instances of self-calibration, such as blind deconvolution, have already been around 
for some time. In this work we have taken a step toward building a systematic mathematical framework for 
self-calibration. There remain more open questions than answers at this stage. Some obvious extensions
of the results in this paper will be to consider the case where both $\bx$ and $\bh$ are sparse.
Another useful generalization is the scenario where $\bx$ is replaced by a matrix, like in the direction-of-arrival
problem with multiple snapshots. Lifting the self-calibration problem in this case would lead to a tensor-valued
(instead of a matrix-valued) underdetermined linear system~\cite{FS14}. The computational complexity for solving the associated
relaxed convex problem will likely be too expensive for practical purposes and other methods will be needed.
Extensions of the Wirtinger-Flow approach in~\cite{CLS14} seem to be a promising avenue to derive computationally
efficient algorithms for this and other self-calibration scenarios. Another important generalization is the case when there is mutual coupling between sensors,
in which case $\BD$ is no longer a diagonal matrix. Yet another very relevant setting consists of $\BD$ depending non-linearly on $\bh$, 
as is often the case when the sensors are displaced relative to their assumed locations. In conclusion, we hope that it has become evident that self-calibration 
is on the one hand an inevitable development in sensor design in order to overcome some serious current limitations, and on the other hand it is
a rich source of challenging mathematical problems.

\section*{Acknowledgement}

T.S.\ wants to thank Benjamin Friedlander from UC Santa Cruz for introducing him to the topic of self-calibration.
The authors acknowledge generous support from the NSF via grant DTRA-DMS 1042939.


\end{document}